# Modeling the Regional Effects of Climate Change on Future Urban Ozone Air Quality in Tehran, Iran

Ehsan Mosadegh [1, *], Khosro Ashrafi [2], Majid Shafiepour Motlagh [3], and Iman Babaeian [4]

[1] Atmospheric Sciences Graduate Program, University of Nevada, Reno, NV 89557, USA.; emosadegh@nevada.unr.edu
[2] Department of Environmental Engineering, Graduate Faculty of Environment, University of Tehran, Tehran, Iran.; khashrafi@ut.ac.ir
[3] Department of Environmental Engineering, Graduate Faculty of Environment, University of Tehran, Tehran, Iran.; shafiepourm@yahoo.com
[4] Climate Research Institute, Atmospheric Science and Meteorological Research Center, Mashhad, Iran; i.babaeian@gmail.com
* Corresponding author: emosadegh@nevada.unr.edu

**Abstract**

Quantifying the impact of climate change on future air quality is challenging in air quality studies. To investigate to what extent climate change can affect future summertime Ozone ($O_3$) air quality in an urban area, the number of future exceedances of $O_3$ air quality standards associated with regional climate change is projected by developing a statistical approach.
An artificial neural network model is employed to simulate hourly $O_3$ concentrations. The model is developed based on hourly observed values of temperature, solar radiation, nitrogen monoxide and nitrogen dioxide which are monitored during summers (June, July and August) of 2009–2012 at an urban air quality station in Tehran, Iran. Climate projections by HadCM3 GCM over the study area, driven by IPCC SRES A1B, A2 and B1 emission scenarios, are downscaled by LARS-WG5 model over the periods of 2015–2039 and 2040–2064. Further calculations are then performed to obtain future hourly temperature and solar radiation values by developing the future diurnal patterns of the variables in the study area. Future hourly $O_3$ concentrations are then projected for five summers in each climate period. The projections are calculated by assuming that current emissions conditions of $O_3$ precursors remain constant in the future. Therefore, only the impact of climate change on future $O_3$ concentrations is investigated in this study. The employed $O_3$ metrics include the number of days exceeding one-hour (1-hr) (120 ppb) and eight-hour (8-hr) (75 ppb) $O_3$ standards and the number of days exceeding 8-hr Air Quality Index (AQI) in the study area.
GCM simulations indicate that the surface temperature in the study area increases over all months of the year. The projected increases in solar radiation and decreases in precipitation in future summers along with summertime daily maximum temperature rise of about 1.2 °C and 3 °C in the first and second climate periods respectively are some indications of more favorable condition for $O_3$ formation over the study area in the future. Based on pollution conditions of the violation-free summer of 2012, the summertime exceedance days of 8-hr $O_3$ standard are projected to increase in the future by about 4.2 days in the short term and about 12.3 days in the mid-term. Similarly, based on pollution conditions of the polluted summer of 2010 with 58 $O_3$ exceedance days, this metric is projected to increase about 4.5 days in the short term and about 14.1 days in the mid-term. Moreover, the number of *Unhealthy* and *Very Unhealthy* days in 8-hr AQI is also projected to increase based on pollution conditions of the both summers.

Ozone, climate change, air quality modeling, artificial neural networks, statistical downscaling, Tehran

1. Introduction

Intergovernmental Panel On Climate Change (IPCC) projections indicate that climate change may influence future air quality and the magnitude of the impact varies from one region to another (IPCC, 2007). One of the challenges associated with air quality studies is to quantify this influence on air pollutants such as Ozone (O3) and PM which are sensitive to climate changes (Jacob and Winner, 2009). Surface O3, which is one of the most important air pollutants, degrades public health by damaging the respiratory system. It is a secondary pollutant which means it is not emitted from a particular source, but is produced through complex photochemical reactions among its biogenic and anthropogenic precursors such as NOx, NMVOC, CO and CH4 in the presence of high temperature and abundant sunlight (Jacob and Winner, 2009; Seinfeld and Pandis, 2006; Steiner et al., 2006). NOx and CO come from combustion sources, but NMVOC and CH4 have a number of natural and anthropogenic sources (Guenther et al., 2000; Sillman, 1999). Therefore, due to its photochemical nature, O3 concentrations generally peak during the summer season when meteorological conditions are often favorable for its formation. O3 has an atmospheric lifetime of about few days in the boundary layer with global sinks of dry deposition and photolysis in the presence of water vapor (Jacob and Winner, 2009). This oxidant pollutant irritates pulmonary system and decreases lung function. O3 is believed to be associated with premature mortality and exposure to its elevated concentrations irritates people who have respiratory diseases such as asthma and pneumonia (Bell et al., 2007, 2004; Ebi and McGregor, 2008; Gryparis et al., 2004; Ito et al., 2005; Mudway and Kelly, 2000).

High precursor emissions combined with favorable meteorology can lead to formation of high O3 concentrations in the lower troposphere. The simplified process of the NO2 - O3 photolytic cycle in the atmosphere is shown in Figure 1. Photolysis of NO2 is one of the main reactions which result in producing the majority of O3 in troposphere (Steiner et al., 2006). NO2 absorbs radiation energy from sun and splits NO2 molecules into NO molecules and highly reactive oxygen atoms (O). Oxygen atoms immediately react with oxygen (O2) in the atmosphere to produce O3. NO is highly reactive and in the presence of other atmospheric constituents such as CO, CH4, and NMVOCs engages in a set of reactions which finally oxide NO to NO2 before NO consumes the produced O3. Thus, since NO molecules do not adequately exist to react with O3 molecules, O3 accumulates in the atmosphere.

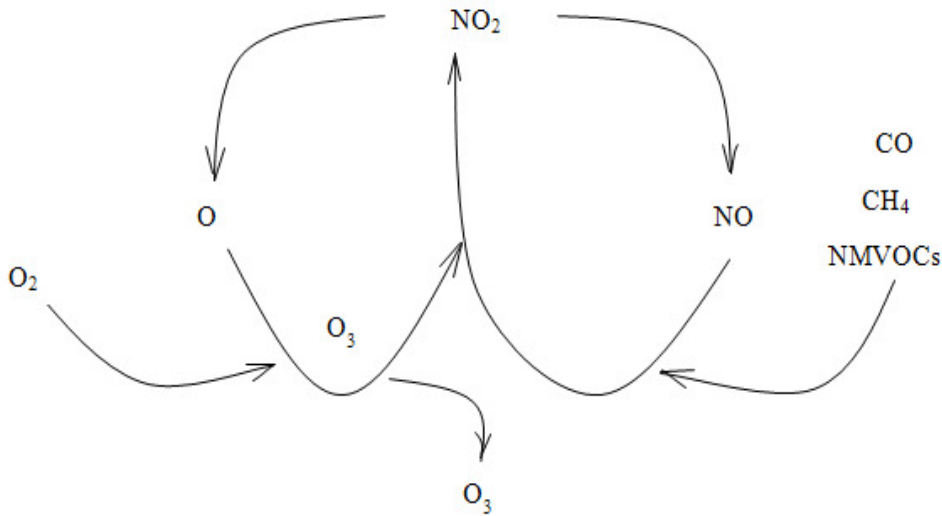

Figure 1. The photolytic cycle of NO2-O3 in the atmosphere.

Meteorological parameters play an important role in O3 production. Temperature, solar radiation, atmospheric moisture, wind, mixing height, precipitation and cloud cover are identified to be correlated with O3 (Camalier et al., 2007; Dawson et al., 2007; Leibensperger et al., 2008; Mott et al., 2005; Nilsson et al., 2001a, 2001b; Ordóñez et al., 2005; Wise and Comrie, 2005). Among these variables, O3 is highly sensitive to temperature (Cox and Chu, 1996; Dawson et al., 2007; Jacob et al., 1993; Sillman and Samson, 1995). The emission of biogenic VOCs which is a temperature dependent process can produce a considerable amount of O3 in high temperatures (Fuentes et al., 2000; Lee and Wang, 2006; Narumi et al., 2009). However, in addition to temperature, solar radiation is also necessary for the photochemical process of O3 formation in the atmosphere. The correlation between these variables is significant especially in summers when high radiation and temperature result in summertime high O3 concentrations (Ordóñez et al., 2005).

Several studies have investigated the potential impact of climate change on future O3 formation. Two major sources of uncertainty in these projections are estimating the future emissions of O3 precursors and projecting the meteorological factors that strongly influence air quality (Dawson et al., 2007; Ebi and McGregor, 2008; Steiner et al., 2006).

Some studies have investigated the influence of projected changes in climate variables on future O3 concentrations by assuming no changes in the emissions of O3 precursors (Dawson et al., 2009; C Hogrefe et al., 2004; Liao et al., 2006; Mickley et al., 2004; Murazaki and Hess, 2006; Racherla and Adams, 2006; Sousounis et al., 2002). The results of these and other similar studies indicate that the projected changes in climate variables are expected to increase future O3 concentration levels over and near polluted regions. The extent of this increase, although varies in different regions, highlights the role of future meteorology conditions in O3 production and suggests that future meteorological parameters will shift toward more favorable conditions for O3 formation. For example Murazaki and Hess (Murazaki and Hess, 2006) used MOZART-2 chemical transport model and SRES A1 scenario, and simulated surface O3 levels for two periods of 1990–2000 and 2090–2100 by considering climate change alone. By comparing exceedances of maximum daily 8-hr average in the two periods, they projected that the number of days exceeding the 80 ppb will increase up to 12 additional days annually by the end of the century. This increase is expected due to the projected changes in various climatic factors such as temperature, water vapor, cloud cover, ventilation conditions, and lightning NOx, although no considerable changes in precipitation patters or planetary boundary layer height were distinguished in the projections. In another study, Liao et al. (Liao et al., 2006) by using a tropospheric Chemistry-Aerosol model and simulating the climate in 2100 under SRES A2 scenario studied the effect of climate changes on future tropospheric chemistry. Their projections indicated that the global O3 burden would decrease at the end of the century due to warmer climate and faster removal of O3. However, the surface layer O3 concentrations would increase in polluted regions as a result of the increased stagnation, the elevated emissions of biogenic VOCs, the increased temperature-induced PAN decomposition and the increased O3 production due to the increased water vapor at the presence of high NOx levels.

Some sensitivity studies have also accounted the influence of changes in emissions of the O3 precursors together with climate change to analyze future sensitivity of the study regions to O3 productions under the impact of climate change (Dawson et al., 2007; C. Hogrefe et al., 2004; Millstein and Harley, 2009; Orru et al., 2013; Steiner et al., 2006). For example, Hogrefe et al. (C. Hogrefe et al., 2004) in a sensitivity analysis studied the changes in O3 concentration in 5 summers of 2050s relative to 1990s and found that after the changed boundary conditions, regional climate change would have the largest effect on summertime average daily maximum 8-hr O3 concentrations. Moreover, changes in the fourth highest summertime 8-hr O3 concentration were more influenced by changes in regional climate than other analyzed factors. They concluded that regional climate change should be considered as important in attaining future air quality standards as changes in anthropogenic O3 precursor emissions and boundary conditions. In another sensitivity study, Steiner et al. (Steiner et al., 2006) evaluated the impact of changes in O3-related climate variables and emissions on O3 production for the central California in 2050 by using a high resolution chemical transport model. They found that combined climate perturbations (such as increases in temperature and water vapor together with temperature-induced increase in biogenic VOC emissions) yield to increased peak O3 concentrations. However, the simulations of combined climate change and emission reductions indicate that future emission reductions could partly offset the future increases of O3 concentrations due to climate change alone. Their results indicate that sensitivity of O3 to climate change is regionally different and the sensitive regions may experience more exceedances despite the present emission reduction policies and therefore additional control on pollution emission reductions will be needed.

In a comprehensive sensitivity study, Dawson et al. (Dawson et al., 2007) investigated the sensitivity of O3 production to climate change over the eastern USA by investigating the influence of several meteorology factors on O3 exceedances of a base case O3 episode in the middle of July 2001. The sensitivity of O3 concentrations to temperature, wind speed, absolute humidity, mixing height, cloud liquid water content and optical depth, cloudy area, precipitation rate, and precipitating area extent were investigated individually by perturbing each variable to varying degrees over the study area. They observed that temperature had the largest influence on both peak 8-hr O3 and standard exceedances among other variables due to the increased decomposition of PAN to O3 precursors. Absolute humidity, mixing height and wind speed, despite having various relative effects, contributed to changes in O3 concentrations. Increased water vapor led to increases in summertime peak 8-hr O3 concentrations in polluted areas while it decreased peak 8-hr O3 concentrations in remote areas. However, increase in wind speed and mixing height led to decrease in O3 concentrations due to more ventilation conditions. The results of this study are consistent with other sensitivity studies which highlight the important role of meteorology in O3 air quality (Baertsch-Ritter et al., 2004; Mickley et al., 2004; Murazaki and Hess, 2006; Racherla and Adams, 2006).

Few studies have addressed the impact of climate change on future O3 air quality by employing statistical approaches (Holloway et al., 2008; Varotsos et al., 2013; Wise, 2009). Constant local O3-meteorology relationship and constant emissions based on current conditions are the restrictive assumptions of these approaches. For instance, Holloway et al. (Holloway et al., 2008) used a statistical approach to investigate the impact of global climate change on summertime O3 mixing ratio over Chicago area during the next 100 years by employing three GCMs driven by A1FI and B1 SRES emission scenarios. They drew a statistical relationship between observed O3 concentrations and meteorological variables which consisted of temperature, solar flux and horizontal surface winds. By employing a statistical technique to downscale the GCM data and assuming that the relationship between observed meteorology and O3 would remain constant in the future, they projected the potential impact of future meteorology on O3 exceedances of 84 ppb. Their projections showed consistency with the results from dynamical studies over the study area. However, the projections were only based on the impact of global climate change alone and neither changes in emissions of O3 precursors nor changes in local meteorology were considered in the simulations. Similarly, Varotsos et al. (Varotsos et al., 2013) investigated the impact of climate change on future O3 air quality in Europe over the periods of 2021–2050 and 2071–2100 under the SRES A1B scenario. They developed a statistical relationship between daily maximum temperature and hourly O3 concentrations and investigated the impact of climate change on the number of days with O3 exceedances of 60 ppb. They observed that higher daily temperatures due to climate change will result in considerable increases in O3 exceedance days in the future. Their results agreed with the results of GISS/GEOS-CHEM dynamical model over most sub-regions in the study area. However, similar to Holloway et al. (Holloway et al., 2008), this study was conducted under the assumptions that current emissions levels and present relationship of O3-temperature would remain constant in the future.

Dynamical models have distinct advantages over statistical approaches. One of the main disadvantages of the statistical approaches is that physical and chemical processes involved in O3 formation such as photochemical interactions between O3 and its precursors or the relationship between O3 and emissions of its precursors cannot be incorporated in the modeling process (Varotsos et al., 2013). However, some benefits of statistical models cannot be ignored. statistical models are widely known for their computationally inexpensive cost and capability of rapid climate change impact assessment by employing various climate models and scenarios (Varotsos et al., 2013). On the other hand, due to the coarse spatial resolution of GCM models, some of small-scale but important processes are not captured in the simulations (Holloway et al., 2008). Furthermore, dynamical models that are developed based on current physical parameterizations may not be able to perfectly simulate future changes in climate variables. For instance, Lynn et al. (Lynn et al., 2004) showed that in order for climate change simulations to provide a realistic estimate of changes in temperature, models should correctly simulate the diurnal precipitation over the study region. In their study, they showed that the patterns were differently simulated in regional and global climate models.

Most of the climate change assessment studies have usually been conducted on regional and global scales. Mostly, they have only projected future O3 concentrations in domain-average responses to climate changes over the study regions. Few studies have addressed changes on an urban scale by considering health related indices. Furthermore, most statistical studies have only taken into account the relationship between O3 and future meteorology. However, Steiner et al. (Steiner et al., 2010) suggested that projecting future O3 air quality is imperfect without considering the influence of O3 precursors together with meteorological variables in the assessment process.

Climate change projections (IPCC, 2021) indicate that that projected changes in climate variables such as precipitation and temperature (Mosadegh and Babaeian, 2022a) will impact different components of the climate system with different magnitude and confidence and in all regions of the world (Mejia et al., 2018; Mosadegh et al., 2018; Mosadegh and Nolin, 2020). Several studies have addressed the issue of air quality in Tehran (Arhami et al., 2013; Ashrafi, 2012; Atash, 2007; Halek et al., 2004; Hosseinpoor et al., 2005; Hoveidi et al., 2013). However, a few studies have investigated uncertainty of those climate projections over the 21st century (Mosadegh and Babaeian, 2022b), and the extent that the projected climate variables can affect air pollution of Tehran region (Mosadegh, 2013). The present study is the first attempt to evaluate the regional impact of climate change on air quality in Iran. The aim of this study is to develop and apply a statistical approach to investigate the impact of climate change on future O3 air quality on a local scale in an urban environment. In this study, an artificial neural network was used as a predictive tool which is capable of capturing nonlinearities in atmospheric processes such as O3 formations (Comrie, 1997; Gardner and Dorling, 1998). The projected O3 concentrations were analyzed based on exceedances of O3 air quality standards and health-related air quality indices. To simplify the impact assessment process, only climate variables of solar radiation and temperature together with pollutants of NO and NO2 were considered in the simulation process. In this study, the relationship between O3 and local meteorology was partially accounted by considering hourly temperature and solar radiation values in the development process of the air quality forecasting (Artificial Neural Network) model. Emissions of O3 precursors were also taken into account, but were considered constant based on current conditions. Therefore, only the impact of climate change was investigated on future O3 concentrations in Tehran.

2. Methodology

Figure 2 illustrates the principal components of this study which include four major steps: The first step is to develop an air quality forecasting model. This requires the identification of affective variables in the system under study, calibration and test of model performance. The second step is to downscale GCM data under different emission scenarios. In this step, LARS-WG model is used for statistical downscaling of GCM data. Initially, the model is calibrated based on present day climate observations and then it generates downscaled daily climate variables for future periods based on selection of a GCM model and a greenhouse gas emission scenario. The third step is to downscale the daily variables obtained from LARS-WG in the second step from daily scale to sub-daily (hourly) scale. The final step is to develop input scenarios for the air quality forecasting model and to assess the impact of climate change on future O3 air quality.

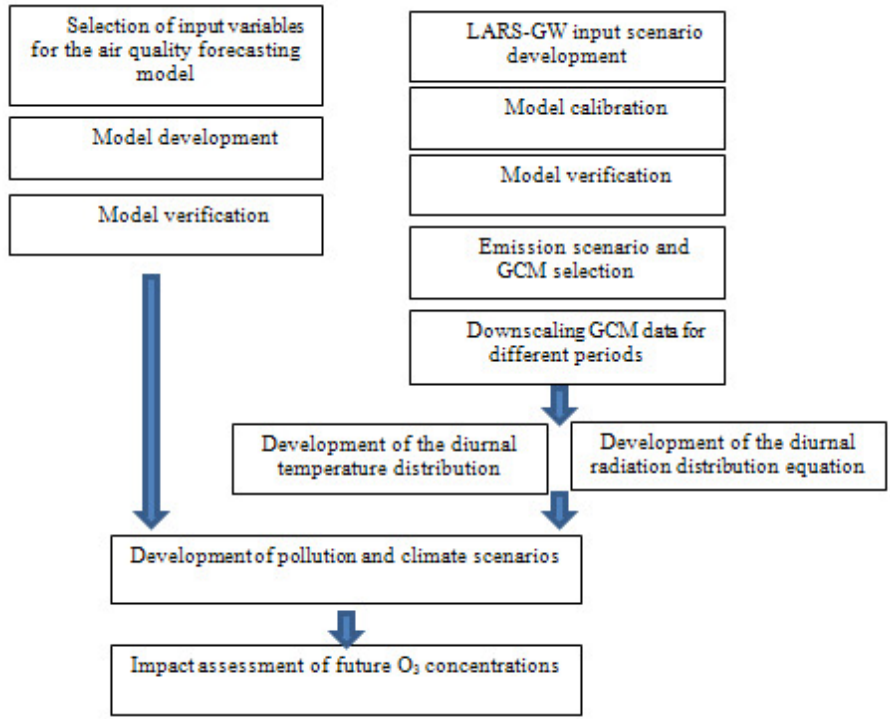

Figure 2. Flowchart of study.

## 2.1. Case study and data

Tehran, the capital of Iran, is the largest city in Iran with the population of more than 10 million people and with the area of approximately 570 square kilometers. The city is located between $35°\ 34' - 35°\ 50'$ N and $51°\ 02' - 51°\ 36'$ E. Tehran is surrounded by mountains to the north and the east, and the wind directions are from the west and the south. Tehran faces serious air pollution problems which stem from several interactive factors such as high population, meteorological conditions and geographical position. Motor vehicles are considered as one of the major sources of air pollution in Tehran urban area due to their high emission of major pollutants such as CO, PM10 and NO2 (Halek et al., 2004). In Tehran, air pollution concentrations are monitored by Air Quality Control Company (AQCC) and Department of Environment (DOE) in several air quality stations.

In this study, the air quality data were obtained from the AQCC Golbarg air quality monitoring station in the east of Tehran at $35°\ 43'$ N and $51°\ 30'$ E. In order to develop and evaluate the air quality forecasting model (ANN), hourly monitored variables were used which consist of Nitrogen Monoxide (NO), Nitrogen Dioxide (NO2), O3, Solar Radiation (SR) and Temperature (T) collected at this station during 2009–2012. Meteorological data were obtained from the *Dushan Teppeh* station, the nearest synoptic station located at $35°\ 42'$ N and $51°\ 20'$ E with the height of 1209.2 m above sea level. To calibrate the LARS-WG statistical downscaling model, meteorological variables which consist of daily minimum temperature, maximum temperature, total precipitation and total sunshine hours during 1972–2009 were used.

## 2.2. Statistical downscaling with LARS-WG

Despite the significant increase in the resolution of General Circulation Models (GCMs), they cannot yet predict the meteorological outputs for small scales such as a city. Different dynamic and statistical models have been developed to downscale the GCM outputs (Wilby et al., 2004). Stochastic weather generators (WG) are one of the statistical downscaling tools which generate daily time series of climate variables (Semenov, 2007; Wilks and Wilby, 1999). In this study, Long Ashton Research Stochastic Weather Generator (LARS-WG) is employed to downscale the GCM projections and to estimate future changes in temperatures, solar radiation and precipitation over the study area.

LARS-WG (Semenov and Barrow, 2002) is a stochastic weather generator (WG). The model takes observed daily minimum temperature, maximum temperature, total precipitation and total sunshine hours as its inputs and generates synthetic daily time series at any local scale. LARS-WG generates local-scale climate change scenarios for a given site by adjusting baseline parameters, calculated from baseline observed weather at the site, with projected GCM Δ-changes, calculated based on an SRES emission scenario and a future climate period, for each climatic variable (Semenov and Stratonovitch, 2010).

In LARS-WG, the production of synthesis daily time series comprises three steps: model calibration, quality tests and daily time series generation. In the calibration step, statistical characteristics of the climatic variables are computed from the probability distributions of observed daily weather data at a given site (Semenov, 2007). This is followed by finding some semi-empirical distributions of observed data, such as probability distributions of the wet and dry durations. In the quality test step, the statistical characteristics of the observed and simulated weather data are analyzed to determine if any statistically-significant differences exists between them. A number of statistical tests are used in this comparison to evaluate the ability of LARS-WG to reproduce the baseline climate: Student's t-test and F-test are two parametric tests that are respectively used for comparing means and standard deviations of the observed and simulated data sets. Kolmogorov-Smirnov (K-S) test, which is a non-parametric test, is also used for comparing probability distributions of observed and generated daily time series. These tests are based upon the assumption that the observed weather is a sample from the true climate at the site. The tests evaluate the statistical differences between generated and observed data. Each test computes a p-value which examines the probability that both data sets have similar distributions. Low p-values, especially below a selected significance level, denote that generated climate is not statistically similar to the true climate at the site (Semenov and Barrow, 2002). In the final step, by applying projected climate change scenarios of a desired GCM, LARS-WG generates synthetic daily time series which have the same statistical characteristics as the observed daily time series of the climatic variables.

In this study, in addition to the mentioned statistical tests such as t- and K-S test, the ability of the LARS-WG to simulate the baseline climate variables at the given site was evaluated by calculating the Pierson correlation coefficient (R) and error indices such as mean bias error (MBE), mean absolute error (MAE) and root mean square error (RMSE) given by equations (1-3), respectively

$$MBE = \frac{\sum_{i=1}^{n}(O_i - P_i)}{n} \tag{1}$$

$$MAE = \frac{\sum_{i=1}^{n}|(O_i - P_i)|}{n} \tag{2}$$

$$RMSE = \sqrt{\frac{\sum_{i=1}^{n}(O_i - P_i)^2}{n}} \tag{3}$$

where $O_i$ is the $i$th target data (observed), $P_i$ is the corresponding simulated data (predicted), and $n$ is the total number of evaluated samples.

R indicates the similarity between observed and simulated concentrations. MBE indicates if the model under- or over-estimates the observed concentrations. MAE and RMSE measure the residual error and indicate the difference between the observed and simulated concentrations.

## 2.3. The air quality forecast model

With recent advances in deep learning for pattern recognition, performance of these networks for the task of prediction in different fields of environmental science has progressed even with small amount of training data (Alibak et al., 2022; Nejatishahidin et al., 2022). Application of artificial neural networks (ANN), especially multilayer perceptions (MLP) in the field of air quality has been evaluated in many studies (Chaloulakou et al., 2003; Comrie, 1997; Gardner and Dorling, 2000; Marzban and Stumpf, 1996; Niska et al., 2004; Schlink et al., 2003; Sousa et al., 2007). Application of the neural networks in forecasting O3 concentrations has been compared with other statistical tools such as multivariate linear regression models, and the results indicate that the ANNs especially the MLP neural network has a better performance over other techniques in modeling the O3 nonlinear associations (Gardner and Dorling, 1998). Unlike other statistical techniques, the MLP does not require prior knowledge of the process, especially about the data distribution of input variables. It can accept multiple inputs having different characteristics without considering any assumption regarding the relationship between the variables. Furthermore, It can model highly nonlinear processes by its activation and transfer functions in the hidden layers (Rahnama and Noury, 2008). These features make MLP a suitable tool for modeling complex, nonlinear phenomena such as O3 formation in the atmosphere.

Basic structure of every neural network involves inter-connected nodes that are arranged in layers. The architecture of every neural network is composed of an input layer, one or more hidden layers and an output layer. Each node in each layer is connected to every node in neighboring layers. Every node in the hidden and output layers consists of activation and transfer functions which incorporate the nonlinearities into the modeling process.

Initially, in each node, the activation function value is calculated. Then the calculated value passes through a transfer function which can be linear or nonlinear. This process is identical for all nodes in hidden and output layers. The input layer, however, does not contain any activation or transfer function and serves merely to transfer inputs to the network. Finally, the output of the system is compared with the target value and output error of the modeling system is calculated by

$$E(n) = \frac{1}{2}\sum_{k=1}^{p}[O_k(n) - S_k(n)]^2 \qquad (4)$$

where $O_k(n)$ is the desired target response (observed) and $S_k(n)$ is the actual response (simulated) of the network for the $k$th neuron and the $n$th training pattern. The objective of the training phase is to reduce the output error of the modeling system to its minimum. In back-propagation training algorithm, this is accomplished by distributing the output error back into the system among network weights and adjusting the weights so that the final output error (Eq. (4)) approximates the target value with a selected error goal

In this study, a four-layer feed forward MLP (FFMLP) with a 4-10-10-1 network structure was used (Figure 3). Tangent sigmoid transfer functions (*tansig*) were used in the hidden layers, but a linear transfer function (*purlin*) was used in the output layer. For training the network, the Levenberg-Marquardt back-propagation learning rule (*trainlm*) was used due to its fast speed and accuracy in training the system (Beale et al., 2012).

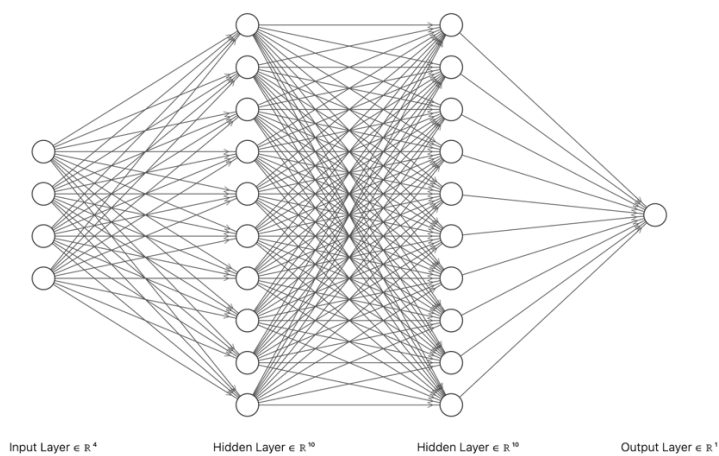

Input Layer ∈ R⁴    Hidden Layer ∈ R¹⁰    Hidden Layer ∈ R¹⁰    Output Layer ∈ R¹

Figure 3. Architecture of the neural network model used in this study.

### 2.3.1. Determination of model inputs

Different methods exist for reducing the number of input variables to an ANN model such as forward selection and backward selection (Noori et al., 2010). However, with regard to several published studies which have addressed the effective variables in O3 production (Ordóñez et al., 2005) and the limited number of available monitored variables in the $Golbarg$ air quality control station, three air quality variables which consist of Nitrogen Oxide (NO), Nitrogen Dioxide (NO2) and O3, and two climatic variables which consist of temperature (T) and Solar Radiation (SR) were selected to develop the air quality forecasting model.

According to several studies, among O3 precursors, Nitrogen oxides (NOx) and Non Methane volatile organic compounds (NMVOCs) are considered as the two main precursors in O3 production in the atmosphere (Jacob and Winner, 2009). Unfortunately, NMVOC concentrations were not monitored in most air quality monitoring stations in the study area, and even in few monitoring stations with observed NMVOC concentrations, the data series were not suitable for developing the air quality forecasting model due to several available defects in the observed time series. Absence of NMVOC concentrations in simulations, which is considered as one of the major limitations of this study, makes the air quality forecasting models simulate O3 concentrations lower than its actual values. Therefore, the forecasting model underestimates the actual values (Liu et al., 1987). It is noteworthy that the selected variables were monitored at $Golbarg$ air quality monitoring stations during the summers (June, July and August) of 2009–2012.

### 2.3.2. Development of the air quality forecast model

Data sets were initially investigated and defective data sets were excluded. The selection of the datasets for developing the forecasting model was limited to the 8 am to 7 pm interval which is the most effective period of O3 production during the day. Finally, about 4000 hourly data sets were obtained for the summers (JJA) of 2009 to 2012 to develop the forecasting model. Then the datasets were shuffled randomly by using a code in MATLAB software to scatter maximum and minimum values evenly over the entire data range. Then, by developing a code in MATLAB software, data sets were divided to three subsets of training, test and validation sets with 60-20-20 percent's of the dataset respectively to avoid the overtraining in the development process.

Since the input and target variables did not have a uniform range of values, a normalization method was used to scale the variables to fall in a certain range and therefore to increase the speed of network training.

In this study, normalization of the variables was performed by the *mapminmax* function in MATLAB to scale the data to the range [-1, 1] before entering the main network. In order to achieve the best relationship between input variables and output target (O3), different network architectures were examined. Finally, the network structure with the smaller error and the higher correlation was selected as the optimal predictive model.

### 2.3.3. Validation of the model based on error indices

In this study, the performance of the forecasting model in simulating the hourly O3 concentrations in both development and validation steps was evaluated by calculating the correlation coefficient (R) and statistical parameters such as mean bias error (MBE), mean absolute error (MAE) and root mean square error (RMSE) similar to the Eqs. (1-3), respectively. After ensuring the accuracy of the simulations and Capability of the developed model to reproduce hourly O3 concentrations with regard to high correlation coefficient and low error indices compared to similar studies (Arhami et al., 2013; Comrie, 1997; Sousa et al., 2007) performance of the forecasting model was assessed based on two performance indices.

### 2.3.4. Validation of the model based on performance indices (PI)

In the present study, the prediction of exceedances of desired O3 air quality concentration thresholds were more important than predicting the exact O3 concentration values. Therefore, to evaluate the accuracy of the developed model in capturing the occasions in which the concentrations exceed a desired O3 air quality concentration threshold, two performance indices of PI1 and PI2 were defined with the following descriptions

PI1: The percentage of correctly identified occasions in which O3 concentrations exceeded a desired threshold.

PI2: The percentage of incorrectly identified occasions.
PI1 indicates the forecasting accuracy of the predicting model at each concentration threshold. This index represents the percentage of the cases that both monitored values and corresponding simulated values exceed a desired concentration threshold and consequently the model is successful in predicting the exceedance. PI2 indicates the overestimation error of the model at each concentration threshold. This index represents the percentage of cases that observations do not exceed the desired concentration threshold, but the model incorrectly indicates that the corresponding simulated values exceed the desired threshold.

In this study, in order to assess the accuracy of the air quality forecasting model in simulating the exceedances, several important O3 concentration thresholds were considered among variousO3 air quality standards and indices. The test dataset of the model was examined to assess the accuracy of the model in predicting the exceedances. The investigated concentration thresholds are significant levels of O3 concentrations in 1-hr O3 air quality standard and air quality index (AQI). Exceeding these threshold concentrations results in occurrence of an *Unhealthy* day (O3 concentration above 125 ppb) and a *Very Unhealthy* day (O3 concentrations above 205 ppb) from AQI perspective, and occurrence of a polluted day (O3 concentrations above 120 ppb) from 1-hr O3 standard perspective. In addition to mentioned thresholds, accuracy of the forecasting model in predicting exceedances of other concentration thresholds (25 ppb and 45 ppb) were also evaluated to enable us to compare the performance of the developed model with similar studies.

### 2.4. Temporal (sub-daily) downscaling

LARS-WG downscales minimum and maximum temperature values for each single day. Solar radiation is also generated in $Mj/m^2.day$ and represents the total solar radiation reaching the earth surface in a single day. However, the air quality forecast model was developed based on hourly (sub-daily) variables and received hourly temperature and radiation values as its inputs.

In order for the LARS-WG output variables to match the air quality forecasting model inputs, LARS-WG outputs were downscaled form daily to hourly (sub-daily) scale by developing the diurnal distribution equations for the temperature and solar radiation at the given site.

### 2.4.1. Calculation of diurnal patterns of future temperature

Accurately estimating the diurnal patterns of temperature in the future is important in assessing the impact of climate change on peak O3 concentration levels (Millstein and Harley, 2009). In this study, to obtain future hourly temperatures, the diurnal pattern of future temperature was anticipated by developing a sinusoidal equation as a function of time of day (Ephrath et al., 1996)

$$T_a = T_{min} + (T_{max} - T_{min}) * S_t \qquad (5)$$

where $T_a$ is the air temperature during day time, $T_{min}$ and $T_{max}$ are the minimum and maximum air temperature respectively and $S_t$ is a function of time $t$, ranging between 0 and 1, which is defined as

$$S_t = sin(\pi \frac{t - LSH + \frac{DL}{2}}{DL + 2P}) \qquad (6)$$

where $DL$ is the day length, $LSH$ is the local time of maximum solar height during the day and $P$ is the delay in the maximum air temperature with respect to the time of maximum solar height at the site.

To estimate the air temperature at night, a declining exponential equation was used (Ephrath et al., 1996)

$$T_a = A + B exp(-\frac{t}{\tau}) \qquad (7)$$

which was developed to

$$T_a = \frac{T_{min(J+1)} - T_s exp(-\frac{\alpha}{\tau} + (T_s - T_{min(J+1)})) exp(\frac{t_a - t_s}{\tau})}{1 - exp(-\frac{\alpha}{\tau})} \qquad (8)$$

where $\tau$ is a time coefficient which was considered 4; $t_s$ and $t_a$ are the time of sunset and the current time, respectively; and $\alpha$ is the night length ($\alpha$ = 24 - DL). Values of $DL$, $LSH$ and $P$ were extracted from temperature and solar radiation graphs which were obtained by studying the variability of parameters during the observation period in the station under study. After inserting these parameters in the equations and by using the daily minimum and maximum temperature from LARS-WG outputs, hourly temperature values were obtained from Eq. (5) and Eq. (8).

### 2.4.2. Calculation of diurnal patterns of future radiation

LARS-WG generates its solar radiation output in $Mj/m^2.day$ as the total daily radiation received by the earth surface in a single day. However, the air quality forecasting model accepts hourly values in $W/m^2$ as its radiation input. In order for the LARS-WG radiation output to match the air quality model input scale, some equation for estimating the diurnal patterns of solar radiation were developed.

The diurnal radiation curve was calculated by obtaining parameters such as daily total radiation ($R_g$), day length ($DL$) and solar elevation ($sin\beta$), computed from the latitude of the site ($L$, radians), the solar declination angle ($\delta$, radians) and time of the day ($t_a$). To compute the sine of the solar elevation ($sin\beta$) some intermediate parameters were needed: $SD$, the seasonal offset of the sine of the solar height

$$SD = sin(L) * sin(\delta) \qquad (9)$$

and $CD$, the amplitude of the sine of the solar height

$$CD = cos(L) * cos(\delta) \qquad (10)$$

The sine of the solar elevation, $sin\beta$, is calculated as

$$sin\beta = SD + CD * cos(\pi \frac{t_a - LSH}{12}) \qquad (11)$$

where $t_a$ is the current time and $LSH$ is the time of maximum solar height. Instantaneous radiation ($R_g$) is computed as

$$R_g = R_g(tot) * sin\beta * \frac{1 + C * sin\beta}{DSBE * 3600} \qquad (12)$$

where $DSBE$ is the daily integral of $sin\beta(1 + sin\beta)$ from sunrise to sunset, calculated as

$$DSBE = arccos(-\frac{SD}{CD}) \frac{24}{\pi} (SD + C * SD^2 + \frac{C * CD^2}{2}) + 12 * CD * (2 + 3C * SD) * \frac{\sqrt{1 - \frac{SD^2}{CD}}}{\pi} \qquad (13)$$

The parameter $C$ (Eqs. (12) and (13)) is a fairly constant meteorological variable, and is considered equal to 0.4 (Spitters et al., 1986). In order to calculate $SD$ and $CD$, a parameter called $\delta$ is used to represent the solar declination angle. For obtaining hourly values of solar declination angle, proposed equations by Jacobson (Jacobson, 2005) was used

$$\delta = arcsin(sin\varepsilon_{ob} * sin\lambda_{ec}) \qquad (14)$$

where $\lambda_{ec}$ represents the ecliptic longitude of the Sun and $\varepsilon_{ob}$ represents the obliquity of the ecliptic. The ecliptic is the mean plane of the orbit of the Earth when it moves around the Sun. The obliquity of the ecliptic represents the angle between the plane of the Earth's Equator and the plane of the ecliptic, which is approximated as

$$\varepsilon_{ob} = 23°.439 - 0°.0000004 N_{JD} \qquad (15)$$

where

$$N_{JD} = 364.5 + (Y - 2001) * 365 + D_L + D_J \qquad (16)$$

$$D_L = \{\left|\frac{(Y-2001)}{4}\right| \quad Y \geq 2001 \ or \ \left|\frac{(Y-2000)}{4} - 1\right| \quad Y < 2001\} \qquad (17)$$

where $N_{JD}$ represents the number of days from the beginning of Julian year 2000. In Eqs. (16) and (17), $Y$ is the current year, $D_L$ is the number of leap days since or before the year 2000 and $D_J$ is the Julian day of the year, which varies from 1 on $1^{st}$ of January to 365 (for non-leap years) or 366 (for leap years) on $31^{st}$ of December. Leap years occur every year evenly divisible by 4. The ecliptic longitude of the Sun is approximately

$$\lambda_{ec} = L_M + 1°.915 sin(g_M) + 0°.020 sin(2g_M) \qquad (18)$$

where

$$L_M = 280°.460 + 0°.9856474 N_{JD} \qquad (19)$$

$$g_M = 357°.528 + 0°.9856003 N_{JD} \qquad (20)$$

$L_M$ and $g_M$ are the mean longitude of the Sun and the mean anomaly of the Sun, respectively. The mean anomaly of the Sun is the angular distance, as seen by the Sun, of the Earth from its perihelion, which is the point in the Earth's orbit at which the Earth is closest to the Sun by assuming that the Earth's orbit is perfectly circular and the Earth is moving at a constant speed.

### 2.5. Development of input scenarios to the air quality forecasting model

Estimating the futureO3 concentrations under climate change required estimating the future pollution emissions and climate conditions for the desired periods which would serve as inputs to the air quality forecasting model (ANN). Therefore, a combination of some pollution and climate conditions were considered as input scenarios to the model to represent some probable future conditions.

#### 2.5.1. Air quality scenarios

Estimating future O3 air quality conditions involve several assumptions and uncertainties (Ebi and McGregor, 2008). Future O3 production is dependent on emissions of its future biogenic and anthropogenic precursors such as NOx and VOCs. Anticipating future emissions of these precursors depend on key factors such as population growth, energy consumption, technology advancement and socio-economic development. Estimation of these key factors involves limitations and uncertainties for the distant future (Millstein and Harley, 2009; Syri et al., 2002; Webster et al., 2002). Furthermore, an artificial neural network is highly sensitive to the resolution of its input predictors. In this study, the air quality forecasting model (ANN) was trained by hourly resolution. Input variables with different resolutions will considerably reduce the accuracy of the forecasted target. Consequently, due to present limitations and uncertainties, we decided to limit our study to only the impact of climate change alone on future O3 air quality. Therefore, current pollution conditions were assumed to remain constant in the future based on hourly monitored NO and NO2 concentrations in the summers of 2010 and 2012, which respectively were considered as representations of highly polluted and unpolluted summertime conditions. The projections based on the two summers can be considered as the high and low ranges of the changes of the number of polluted days under the future climate.

#### 2.5.2. Climate change scenarios

According to the considered assumptions in the Air Quality Scenarios section (3.4.1), we limited our study to only the effect of climate change on current pollution conditions. In order to evaluate this effect, the IPCC B1, A1B and A2 greenhouse gas emission scenarios were used to simulate future climate. These scenarios are among main SRES illustrative marker scenarios which were used by the IPCC in the 4th assessment report and have been the focus of several model inter-comparison studies. In this study, climate projections of HadCM3 AOGCM, developed by Hadley Center, were used to obtain future climate variables. This GCM is a coupled atmospheric-oceanic model which has been used and suggested in several previous studies (Hessami et al., 2008; Holloway et al., 2008; Lioubimtseva and Henebry, 2009; Zarghami et al., 2011). This model simulates the global climate with 19 levels in its atmospheric component with a horizontal resolution of 2.5° by 3.75° degrees (latitude by longitude) and 20 levels in its oceanic component with a horizontal resolution of 1.25° by 1.25° degrees.

### 3. Results and discussion

### 3.1. Verification of LARS-WG

To verify the downscaled results, the ability of LARS-WG to simulate the baseline climate (1972–2009) was evaluated by coefficient of determination ($R^2$), statistical tests such as t-test and K-S test, and statistical parameters such as RMSE, MAE and MBE. Figure 4 demonstrates the linear correlation between the simulated monthly means of the variables by LARS-WG and their corresponding observed values for minimum temperature, maximum temperature and solar radiation in the baseline period. It is noticeable that for every three variables, the correlation is considerably high between the observed and simulated monthly means, which indicates the acceptable ability of LARS-WG to simulate the monthly means in the baseline period.

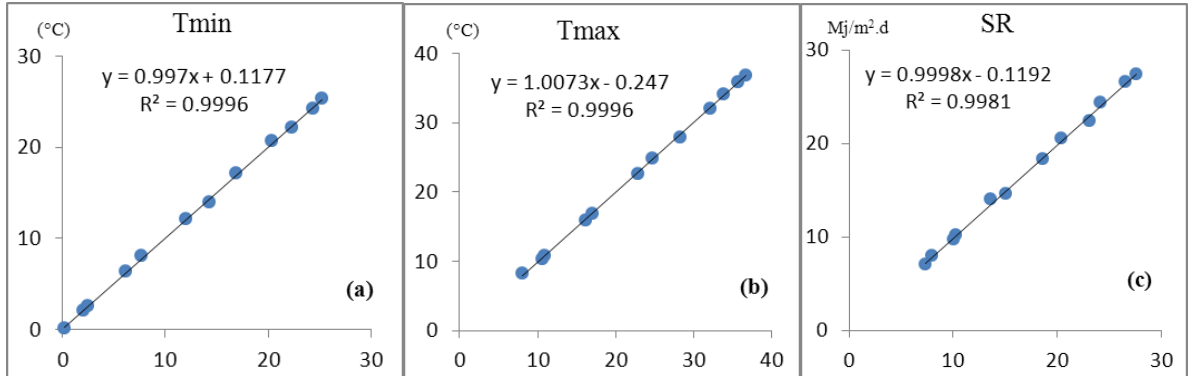

Figure 4. Linear correlation between observed and simulated monthly means of the minimum temperature, maximum temperature and solar radiation by LARS-WG.

Table 1 indicates the calculated statistical parameters for the simulated monthly means of the climatic variables by LARS-WG in the baseline period. Except for precipitation that has the highest simulation error; error indices are relatively low for all variables which demonstrate the acceptable agreement between the observed and simulated monthly means in the baseline period in the study area.

Table 1. Calculated statistical parameters for the simulated monthly means of the variables by LARS-WG in the baseline period (1972–2009)

| Climatic variables | Statistical parameters | | |
| --- | --- | --- | --- |
| | MBE | MAE | RMSE |
| Minimum Temperature | -0.03 | 0.12 | 0.15 |
| Maximum Temperature | 0.08 | 0.19 | 0.23 |
| Solar Radiation | 0.12 | 0.28 | 0.33 |
| Precipitation | 2.9 | -20.2 | -24.5 |

Statistical tests such as t-test and K-S test were also performed for each and every variable to evaluate the ability of LARS-WG to reproduce the probability distributions of daily time series in the baseline climate. Table 2 shows the results of the tests for the solar radiation, precipitation, minimum temperature and maximum temperature. In the mentioned tests, the significance level was set to 0.05. As the Table 2 shows, p-values for all four variables, except for radiation in November, are higher than the 0.05 significance level in all months. This represents the satisfactory performance of the model in simulating daily time series of all variables (except for radiation in November) in the study area.

Table 2. Statistical details of LARS-WG verification

| Month | Precipitation | | | | Minimum Temperature | | | | Maximum Temperature | | | | Solar Radiation | | | |
|---|---|---|---|---|---|---|---|---|---|---|---|---|---|---|---|---|
| | K-S | p-Value | t | p-Value | K-S | p-Value | t | p-Value | K-S | p-Value | t | p-Value | K-S | p-Value | t | p-Value |
| Jan | 0.06 | 1.00 | -0.87 | 0.39 | 0.11 | 1.00 | 0.15 | 0.88 | 0.11 | 1.00 | 0.70 | 0.49 | 0.04 | 1.00 | 0.36 | 0.72 |
| Feb | 0.04 | 1.00 | 0.23 | 0.82 | 0.05 | 1.00 | -0.08 | 0.94 | 0.11 | 1.00 | 0.63 | 0.53 | 0.09 | 1.00 | -0.19 | 0.85 |
| Mar | 0.14 | 0.97 | 1.12 | 0.26 | 0.05 | 1.00 | -0.57 | 0.57 | 0.05 | 1.00 | -0.76 | 0.45 | 0.09 | 1.00 | 0.02 | 0.99 |
| Apr | 0.07 | 1.00 | -0.04 | 0.97 | 0.11 | 1.00 | -0.25 | 0.80 | 0.05 | 1.00 | -0.42 | 0.68 | 0.04 | 1.00 | 0.08 | 0.94 |
| May | 0.08 | 1.00 | 1.20 | 0.24 | 0.05 | 1.00 | -0.91 | 0.37 | 0.05 | 1.00 | -1.05 | 0.30 | 0.09 | 1.00 | -0.57 | 0.57 |
| Jun | 0.13 | 0.98 | 0.33 | 0.74 | 0.11 | 1.00 | 0.76 | 0.45 | 0.11 | 1.00 | 0.98 | 0.33 | 0.04 | 1.00 | -0.03 | 0.98 |
| Jul | 0.12 | 1.00 | -0.18 | 0.86 | 0.11 | 1.00 | -0.67 | 0.50 | 0.11 | 1.00 | 0.60 | 0.55 | 0.04 | 1.00 | -0.60 | 0.55 |
| Aug | 0.31 | 0.19 | 0.26 | 0.80 | 0.11 | 1.00 | 0.41 | 0.68 | 0.11 | 1.00 | 1.03 | 0.31 | 0.13 | 0.98 | 0.55 | 0.59 |
| Sep | 0.29 | 0.24 | -0.47 | 0.64 | 0.05 | 1.00 | -1.19 | 0.24 | 0.05 | 1.00 | -0.81 | 0.42 | 0.09 | 1.00 | 0.78 | 0.44 |
| Oct | 0.06 | 1.00 | -0.96 | 0.34 | 0.11 | 1.00 | 0.86 | 0.39 | 0.11 | 1.00 | 1.51 | 0.14 | 0.04 | 1.00 | -0.55 | 0.58 |
| Nov | 0.09 | 1.00 | -0.98 | 0.33 | 0.05 | 1.00 | -1.51 | 0.13 | 0.05 | 1.00 | -0.89 | 0.38 | 0.04 | 1.00 | -1.99 | 0.05 |
| Dec | 0.04 | 1.00 | 1.41 | 0.16 | 0.11 | 1.00 | -0.40 | 0.69 | 0.05 | 1.00 | -1.03 | 0.31 | 0.13 | 0.98 | -0.71 | 0.48 |

### 3.2. Regional changes in climate

Figure 5 illustrates the HadCM3 projected absolute changes in surface minimum and maximum temperature for Tehran under A2, A1B and B1 emission scenarios. Projections were obtained for the future periods of 2015–2039 (short term) and 2040–2064 (mid-term) relative to the baseline period (1972–2009). Long term monthly means of observed minimum and maximum temperatures in the baseline period are also illustrated on the diagrams to provide an estimate of the future annual temperature patterns in the study area under climate change.

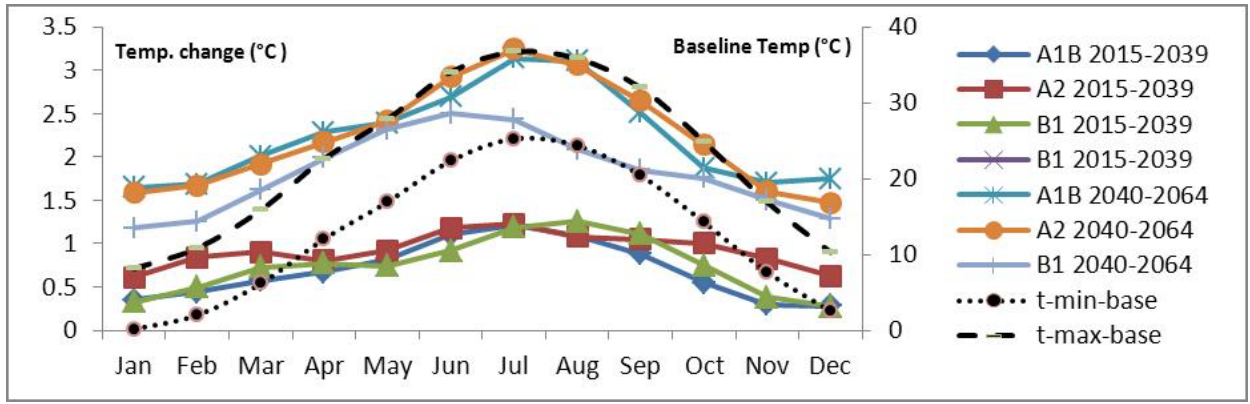

Figure 5. The HadCM3 projected absolute changes in surface minimum and maximum temperature for Tehran for the short term (2015–2039) and mid-term (2040–2064) periods with respect to the baseline (1972–2009) under A1B, A2, and B1 emission scenarios.

HadCM3 projections indicate that the monthly mean surface temperatures are expected to increase under climate change in Tehran. Projections show higher average surface temperatures for all months of the year, but the increase is not uniform throughout the year. Temperature rise is projected to be higher in the warm months (June, July and August), which is an indication of hotter summers in the future. The average surface temperature of the study area is projected to increase by approximately 0.75 °C in short-term and about 2.5 °C in the middle of this century. This temperature rise is expected to exceed 1 °C and 3 °C in the warm month of the year in the short term and mid-term periods respectively. This trend is noticeable in both short term and mid-term climate periods. In the mid-term period, changes in the projections become more distinctive among emission scenarios. Projected changes under A2 and then A1B emission scenarios are expected to be greater than changes under B1 scenario especially in summers where the difference is about 0.5 °C (Figure 5).

Figure 6 depicts the projected relative changes in local precipitation and radiation for the 2015–2039 and 2040–2064 climate periods with respect to the baseline period under the three emission scenarios. Long term monthly means of the variables in the baseline period are also illustrated on the bars to provide an estimate of the future annual patterns of precipitation and radiation in the study area under climate change. Projections illustrate explicit reverse variations in annual patterns of precipitation and radiation under climate change in the future. Projections show that precipitation will decrease in springs and summers, while it will increase in falls and winters with respect to its baseline values. Radiation, in contrast to the precipitation, is projected to increase in springs and especially in summers, and to decrease in falls and winters with respect to its baseline values. The results suggest that maximum decrease in precipitation is expected in summers, about 15% and 30% with respect to the baseline period in short term and mid-term respectively. Unlike the precipitation, the greatest increase in solar radiation is projected in summers, about 1% and 2% in short term and mid-term respectively. A physical interpretation for these reverse patterns is that the decrease in 4 precipitation and cloud cover in summers affects the amount of solar radiation received by the earth surface in the study area.

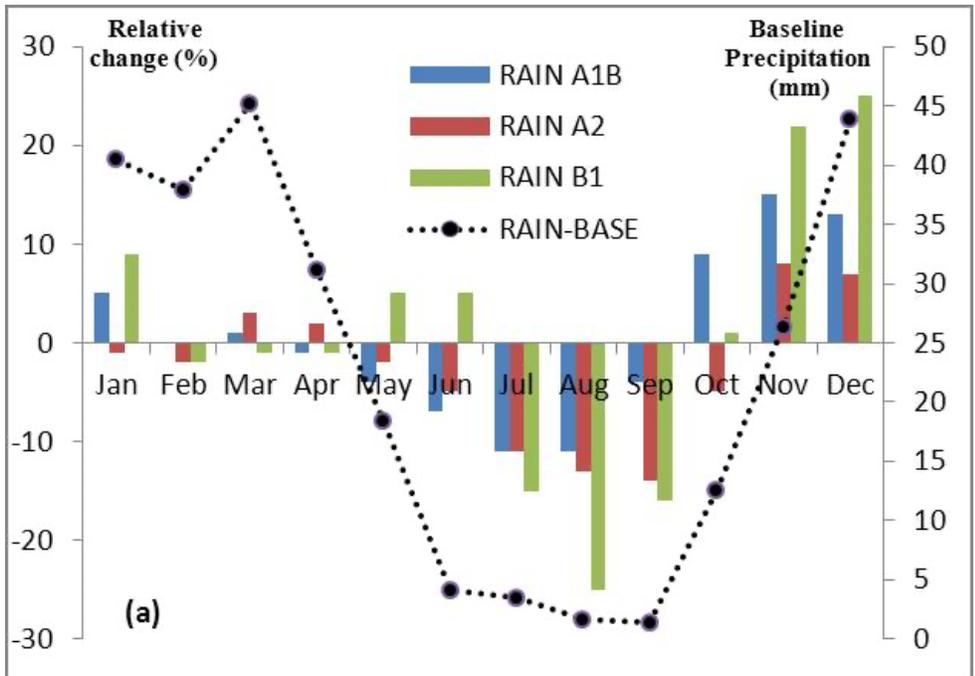
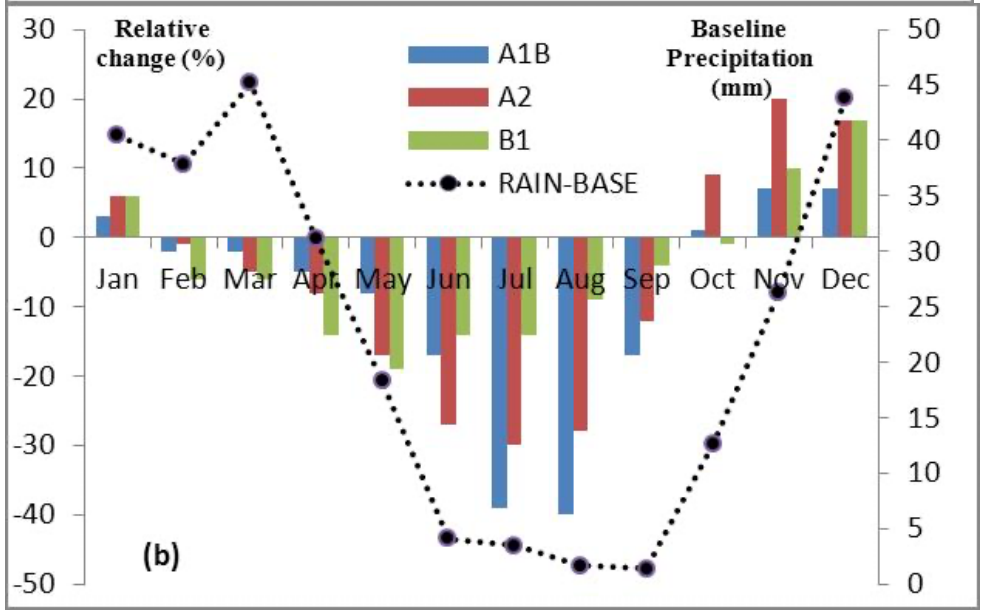

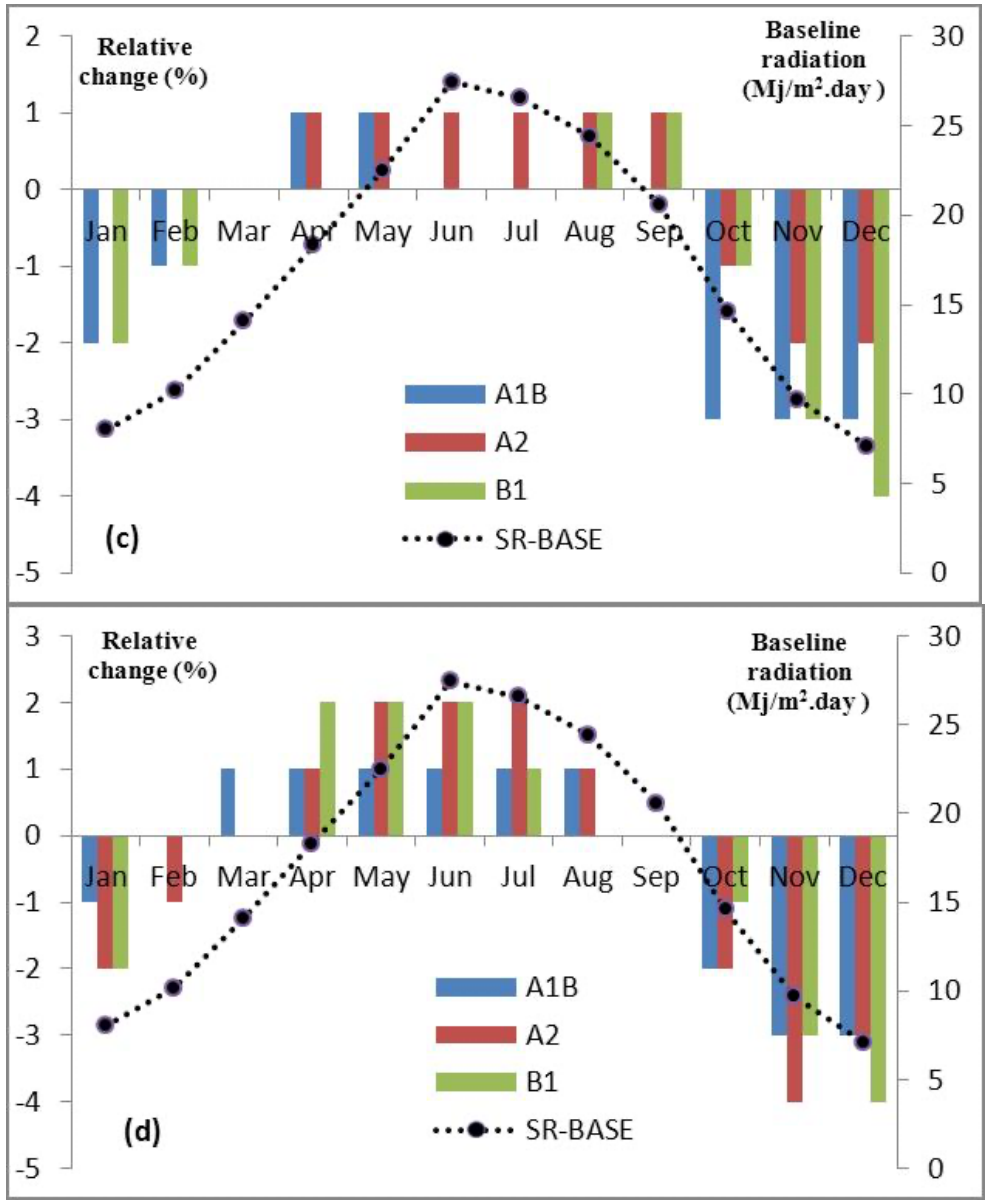

Figure 6. The HadCM3 projected relative changes in precipitation (a and b) and solar radiation (c and d) for Tehran for the short term (2015–2039) (left panel) and mid-term (2040–2064) (right panel) periods with respect to the baseline (1972–2009) under A1B, A2, and B1 emission scenarios.

Climate simulations for future periods over the study area exhibit behaviors favorable to surface $O_3$ formation. In general, HadCM3 GCM model projections show an increase in temperature with the greatest changes in summers under all three emission scenarios. Moreover, solar radiation is projected to increase in summers in all simulations, due to the decreases in precipitation and cloud cover over the study area. These climate change patterns expect to influence $O_3$ production over the city in the future.

### 3.3. Statistical review of air quality in the study area

Figure 7 illustrates the observed monthly means of the air quality variables used in this study that are averaged over the 2009–2012 period. The mean monthly variations of temperature (T) and solar radiation (SR) indicate that the solar radiation in June, and after a month delay, the temperature in July reach their highest values. Therefore, having the highest temperatures and relatively the highest radiation, June, July and August months were considered as warm months of the study area.

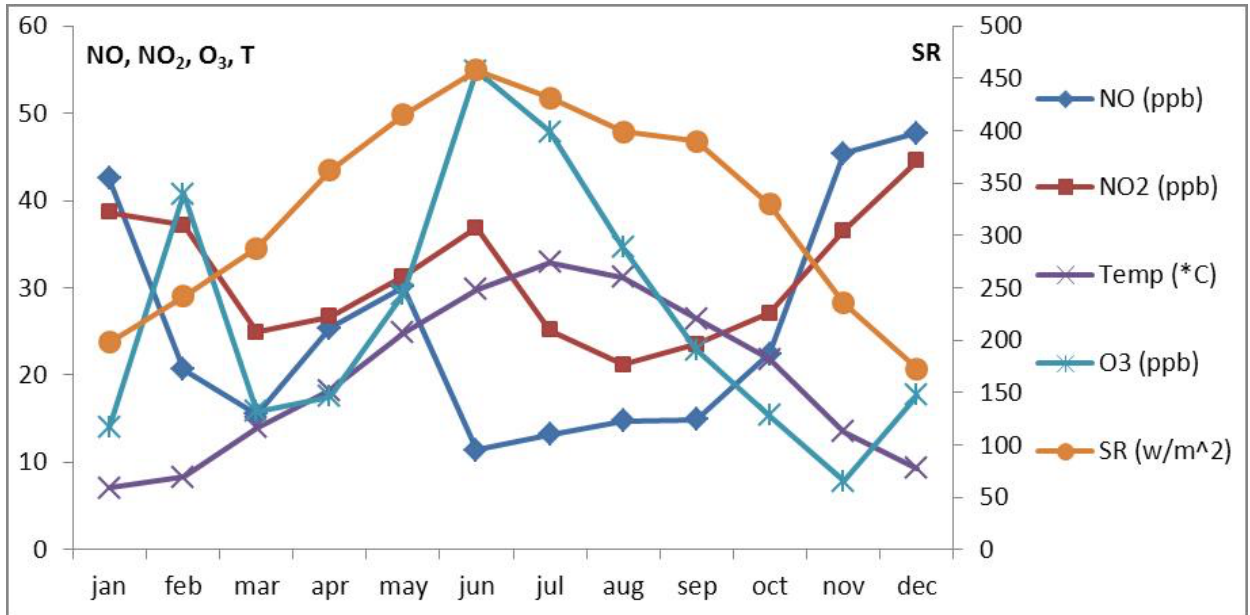

Figure 7. Mean annual cycles of NO, NO2, O3, SR and T at the Golbarg air quality monitoring station for the period 2009–2012.

Moreover, as the Figure 7 indicates, the observed monthly mean O3 concentrations at the Golbarg air quality station have their highest values in the warm months of June, July and August. O3 production in the atmosphere is highly dependent on high temperature which usually occur in the warm months with abundant solar radiation. Also, considering the climate projections over the study area indicate that temperature and radiation will be higher in the warm months in the future, we decided to limit the evaluation of climate change impacts on future O3 concentrations to only the warm months in the study area, i.e. June, July and August.

Among the four summers of 2009 to 2012, summers of 2010 and 2012 had the most and the least number of days with exceedance of O3 air quality standards respectively. The number of exceedance days was much higher in 2010 than in 2012 due to the more favorable meteorological conditions in the summer of 2010.

In the summer of 2010 and in terms of one-hour (1-hr) O3 standard, total of 22 days and in terms of eight-hour (8-hr) O3 standard, total of 58 days exceeded the 120 and 75 ppb concentration threshold respectively, but in the summer of 2012 no polluted day was occurred in terms of any O3 air quality standard. Figure 8 clearly illustrates the difference between the pollution conditions in the two summers. This figure shows the mean diurnal variations of the variables in the summers (JJA) of 2010 (a) and 2012 (b) at the Golbarg air quality station. Table 3 also shows the statistical characteristics of the variables for the two summers. As it is clear, summer of 2010 had higher O3 concentrations in the three months compared to the summer of 2012 in terms of both seasonal means and mean diurnal concentrations. Figure 9 demonstrates the mean diurnal variations of 1-hr and 8-hr O3 concentrations in the two summers of 2010 and 2012 at the Golbarg air quality monitoring station. In the both summers, 1-hr O3 concentrations reached their highest value at around 4 pm, while 8-hr O3 concentrations typically reached their highest value at around noon. Moreover, the difference between the 8-hr O3 concentration levels in the two summers is well demonstrated in Figure 9. The 8-hr O3 concentrations in the summer of 2010 were almost around 70 ppb during the day in June and August, even reached 130 ppb in July, while 8-hr O3 concentrations did not exceed 50 ppb in the summer of 2012.

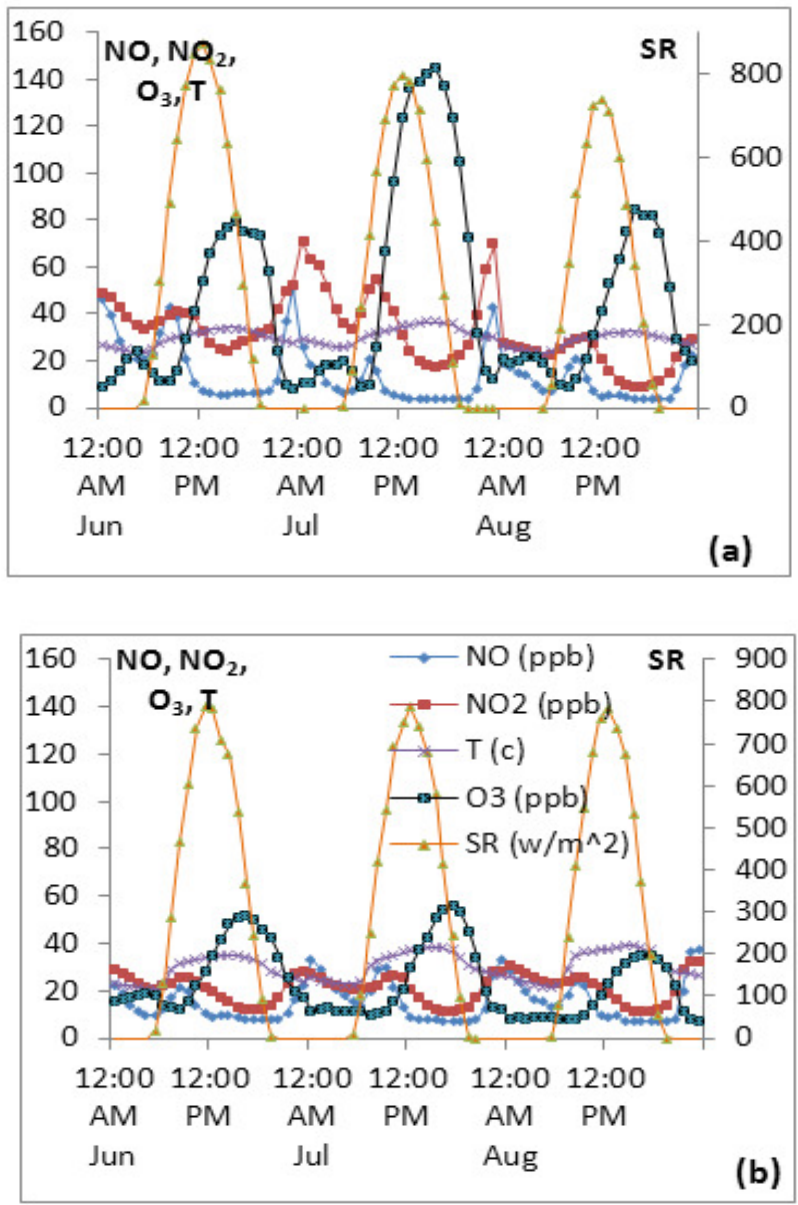

Figure 8. Mean diurnal cycles of NO, NO2, O3, SR and T at the Golbarg air quality monitoring station for the summers of 2010 (a) and 2012 (b).

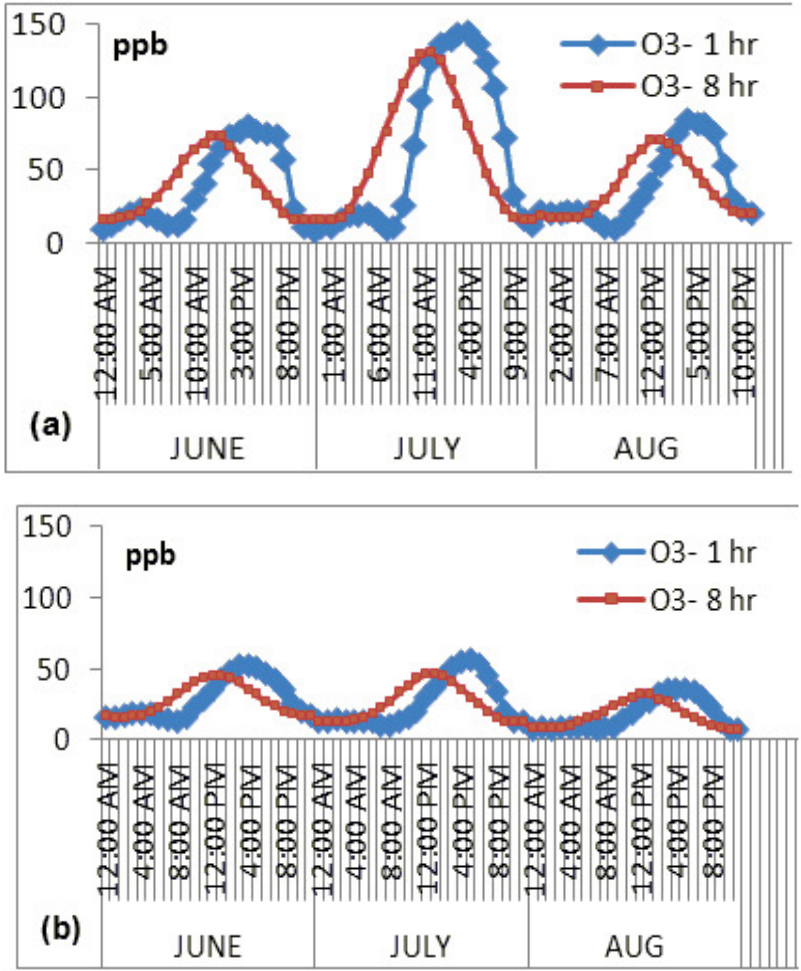

Figure 9. Mean diurnal variations of 1-hr and 8-hr O3 concentrations in the summers of 2010 (a) and 2012 (b) at the Golbarg air quality monitoring station.

Table 3. Statistical review of the pollution and meteorological variables for the summers of 2010 and 2012

| Summer 2010 | Variables | Minimum | Maximum | Average | Standard deviation |
|---|---|---|---|---|---|
| | NO (ppb) | 3.5 | 106 | 9.3 | 13.9 |
| | NO2 (ppb) | 4 | 149 | 26.6 | 17 |
| | O3 (ppb) | 4 | 280.4 | 71.2 | 51.8 |
| | T (C) | 23.58 | 42.67 | 32.6 | 3.1 |
| | SR (w/m2) | 0 | 939 | 506.1 | 270.3 |
| Summer 2012 | Variables | Minimum | Maximum | Average | Standard deviation |
| | NO (ppb) | 6.5 | 82.14 | 11.7 | 9.8 |
| | NO2 (ppb) | 7.2 | 50.17 | 17.5 | 7.26 |
| | O3 (ppb) | 6.5 | 96.53 | 33.45 | 17.56 |
| | T (C) | 20.54 | 43.1 | 35 | 3.66 |
| | SR (w/m2) | 0 | 902 | 493 | 272.28 |

### 3.4. Development and validation of the air quality forecasting model

To find the optimal architecture for the air quality forecasting model, several structures with different numbers of hidden layers and nodes were evaluated. Two out of several various examined architectures with the calculated statistical parameters from the test data sets are shown in Table 4. Statistical parameters indicate that the network with two hidden layers, which has higher correlation coefficient (R) and lower MBE, MAE and RMSE, can better capture the complex and nonlinear relationships among variables of the model. Consequently, the architecture with two hidden layers was selected as the optimal structure for the air quality forecasting model.

Table 4. Calculated statistical parameters for the two developed models

| | No. of neurons | R | MBE | MAE | RMSE |
|---|---|---|---|---|---|
| 1 hidden layer | 10 | 0.812 | -1.77 | 14.5 | 21.38 |
| 2 hidden layers | 10 | 0.84 | -0.9 | 13.8 | 20.43 |

The evaluation criteria of the forecasting model are in the acceptable range compared to other similar studies (Arhami et al., 2013; Comrie, 1997; Sousa et al., 2007). The correlation between the simulated (horizontal axis) and observed O3 concentrations (vertical axis) for the test data sets of the model are shown in Figure 10. The correlation coefficient is about 0.84 which indicates an acceptable agreement between observed and simulated O3 concentrations at the Golbarg air quality monitoring station. The MBE index is about −0.9 ppb. The negative value indicates that the forecasting model underestimates the hourly O3 concentrations about 0.9 ppb under the actual observed values. This can be due to the absence of VOC concentrations in the simulation process (Liu et al., 1987). In comparison with similar studies, the MAE and RMSE, about 13.8 and 20.43 ppb respectively, are also in the acceptable range which indicate the acceptable performance of the air quality forecasting model in predicting hourly O3 concentrations with the least number of input variables (Figure 10).

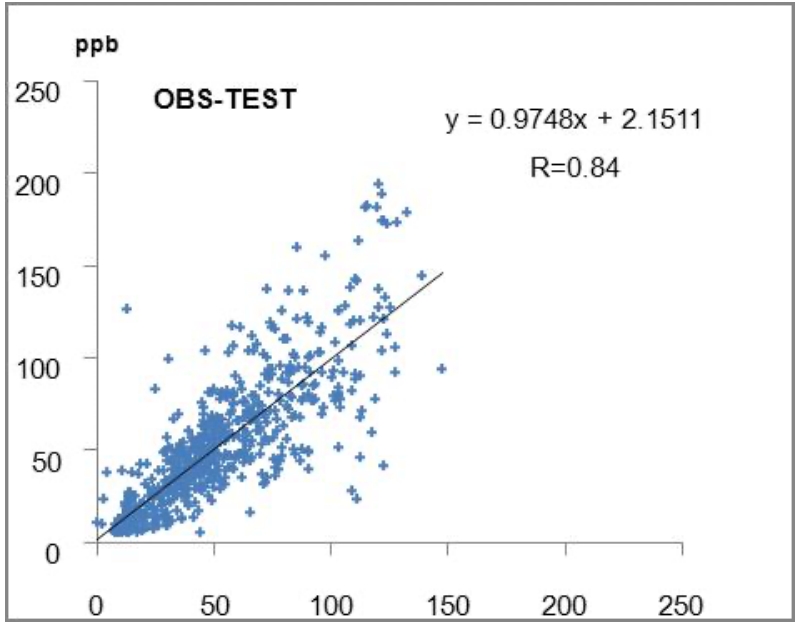

Figure 10. Simulated vs. observed O3 concentrations.

Table 5 illustrates the ability of the air quality forecasting model to capture the exceedances of selected concentration thresholds with their corresponding references. As the table indicates, the developed model gives an acceptable prediction performance compared to the similar study (Nunnari et al., 1998). In detecting exceedances of 25 ppb and 45 ppb thresholds, the model is able to identify 95.7% and 85.9% of exceedances with the overestimation error of 19% and 15.7% respectively that demonstrates a better performance compared to the similar study. Moreover, in higher concentrations (exceedances of 120 ppb) the developed model is able to detect about 40% of exceedances with the overestimation error of about 1%. With regard to the high concentration thresholds and also the number of the model inputs, the forecasting model represents a relatively acceptable performance in predicting the violations.

Table 5. Performance of the forecasting model at selected concentration thresholds with their corresponding references (Results from a similar study are shown in parentheses)

| Reference | Time period | O3 threshold (ppb) | PI 1 (%) | PI 2 (%) |
| --- | --- | --- | --- | --- |
| O3 Standard | 1 hr | 125 | 13.8 | 0.48 |
| O3 AQI | 1 hr | 120 | 39.4 | 0.9 |
| O3 Information Level (EPA) | 1 hr | 90 | 54.5 | 5.4 |
| Nunnari et al. (1998) | 1 hr | 45 | 85.9 (64.57) | 15.7 (4.25) |
| | | 25 | 95.7 (97.75) | 19 (18.03) |

### 3.5. Extracting necessary parameters for calculating future diurnal patterns of temperature and radiation

In section 2.4.1. some equations were developed to obtain the diurnal temperature distribution curve in the study area. In order to develop these sets of equations, *DL*, *LSH* and *P* parameters were needed which were extracted from Figure 8 at the *Golbarg* air quality station.

$LSH$, the maximum solar height, was considered 12 according to Figure 8. $P$, the time lag between the occurrence of maximum temperature and maximum solar height in a day, was considered 3.5 hours according to Figure 8. $DL$, the day length, was obtained from US Navy website (http://www.us.navy.com) according to the location of the station and the study period.

To obtain hourly temperature values during a day, the parameters were replaced in the developed temperature equations (Eq. (8) and Eq. (11)). Then, the downscaled minimum and maximum temperatures from LARS-WG for each day were replaced in the equations and hourly temperatures were obtained for each day.

Solar radiation output from LARS-WG represents daily total radiation received by the earth surface in $Mj/m^2.day$. Downscaled radiation values were distributed during the day according to discussed approach in section 3.3.2 to obtain hourly values in $w/m^2$. In these sets of equations, $LSH$ was considered 12 for the Golbarg air quality station in the study area.

### 3.6. Climate change impacts on Ozone air quality

In this study, the impact of climate change on future O3 concentrations were investigated by considering three Climate projections, A1B (moderate), A2 (warm) and B1 (cool) SRES emission scenarios, and two pollution conditions, summers of 2010 and 2012. The pollution conditions were considered constant based on current conditions and therefore only the impact of climate change on future O3 air quality was investigated by assuming that NO and NO2 levels stay constant based on hourly monitored concentrations in the summers of 2010 and 2012. Finally, six different input scenarios to the air quality forecasting model were obtained and analyzed for the climate periods of 2015–2039 (short term) and 2040–2064 (mid-term).

To assess the changes in summertime O3 concentrations over the two future climate periods, hourly O3 concentrations were calculated for 10 summers (June, July and August) in the future that started from the summer of 2015 with a 5-year interval. To analyze future O3 air quality in the study area, projected concentrations were analyzed from different O3 air quality standards and indices (AQI) which are summarized in Table 6. Moreover, the average number of projected O3 polluted days in each climate period was also calculated and compared with the current conditions' pollution levels.

Table 6. O3 air quality standards and indices used in this study.

| Air quality Index (AQI) | Reference | O3 concentration range (ppb) | Pollution condition |
|---|---|---|---|
| 1 hr | Iran and EPA | 125-204 | Unhealthy |
| | | 205-404 | Very Unhealthy |
| | | 405-504 | Dangerous |
| 8 hr | EPA | 79-115 | Unhealthy |
| | | 116-374 | Very Unhealthy |

| Air Quality Standard | Reference | O3 concentration threshold (ppb) | Pollution condition |
|---|---|---|---|
| 1 hr | Iran and EPA | 120 | Violation of 1 hr standard |
| 8 hr | EPA | 75 | Violation of 8 hr standard |

### 3.6.1. Projected changes in future Ozone air quality standard exceedances

The USEPA considers the 120 and 75 ppb O3 concentrations as the thresholds for violating 1-hr and 8-hr O3 air quality standards respectively. In this study, we used the number of days that hourly O3 concentrations exceeded each of these thresholds, so multiple exceedances within a single day were not counted at the air quality monitoring station.

Figure 11a compares the changes in the number of days that exceed 1-hr O3 standard in the both present pollution conditions. In the summer of 2010, 22 days exceeded 1-hr O3 standard and no exceedances were occurred in the summer of 2012. The projections indicate that the number of polluted days will increase under future climate compared to the base year pollution conditions by assuming no changes in the base year pollution emissions. The number of polluted days grows larger based on pollution conditions in the violation-free summer of 2012, so that in the middle of the century the number of exceedances of 1-hr standard reaches the level of exceedances based on pollution conditions in the highly polluted summer of 2010.

Figure 11b compares the changes in the number of days that exceed 8-hr O3 standard in both present pollution conditions as a result of changes in future climate. Similar to 1-hr exceedances (Figure 11a), the projections show an increase in the number of 8-hr exceedances under the future climate by assuming no changes in the base year pollution conditions. In this graph, due to 8-hr moving average from the projected 1-hr concentrations, variations in the 1-hr diagrams have changed into a smooth diagram which can better illustrate the increase in the number of polluted days in both pollution conditions under the impact of future climate.

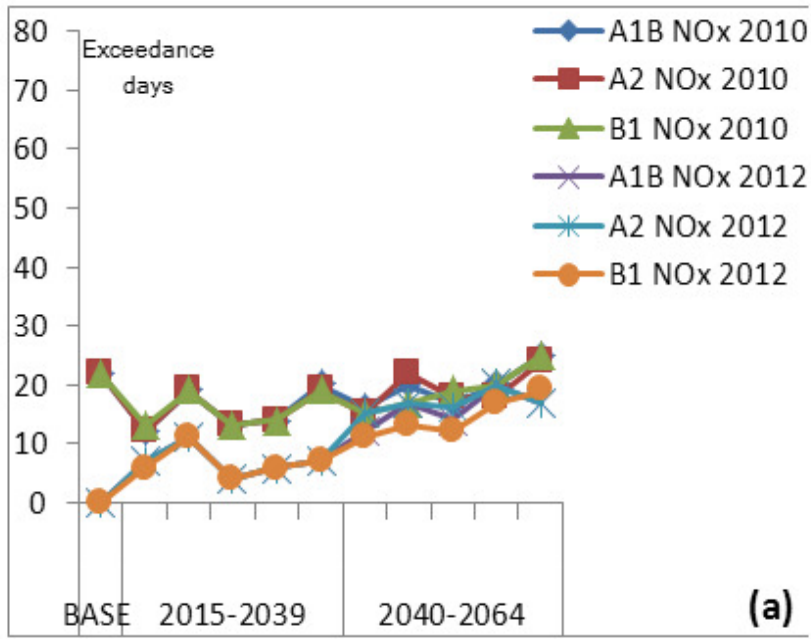

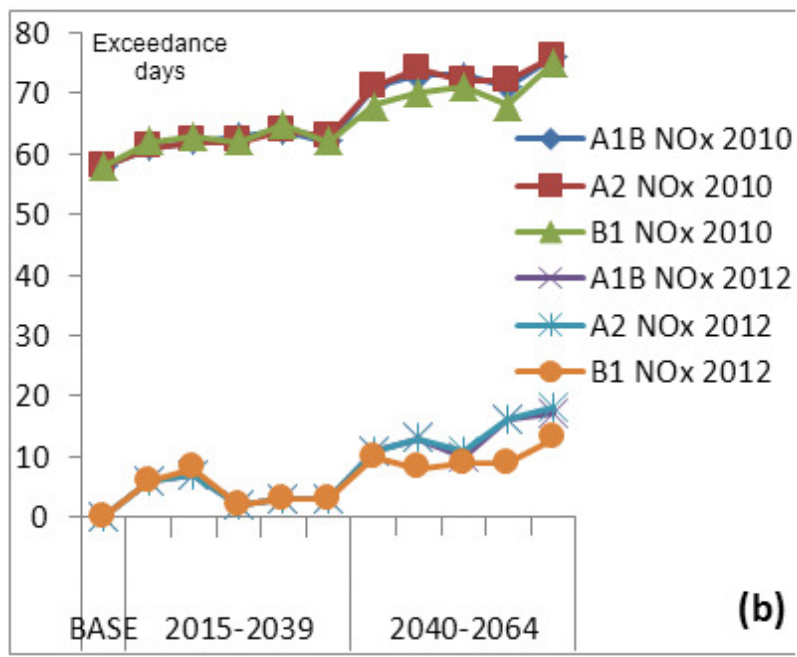

Figure 11. Projected days per summer (JJA) with exceedances of 1-hr (a) and 8-hr (b) O3 standard based on summers of 2010 and 2012.

Figure 12a illustrates the potential impact of climate change on pollution conditions in the summer of 2010. This summer experienced 22 and 58 polluted days that exceeded 1-hr and 8-hr O3 standard respectively. Although the summer of 2010 was a highly polluted summer, climate change still has an increasing influence on the number of projected polluted days. This increase is more notable in changes in 8-hr exceedances.

Figure 12b illustrates the potential impact of climate change on pollution conditions of the summer of 2012. No violation from any O3 standard occurred in the summer of 2012. However, it is expected that polluted days may occur in the future due to changes in climate, and the number of exceedances may increase in future climates. In the short term, projections indicate that changes in climate will mostly impact 1-hr exceedances, but in the mid-term, due to projected higher O3 concentrations, exceedances of 8-hr standard will increase and the controlling standard will shift from 1-hr to 8-hr standard.

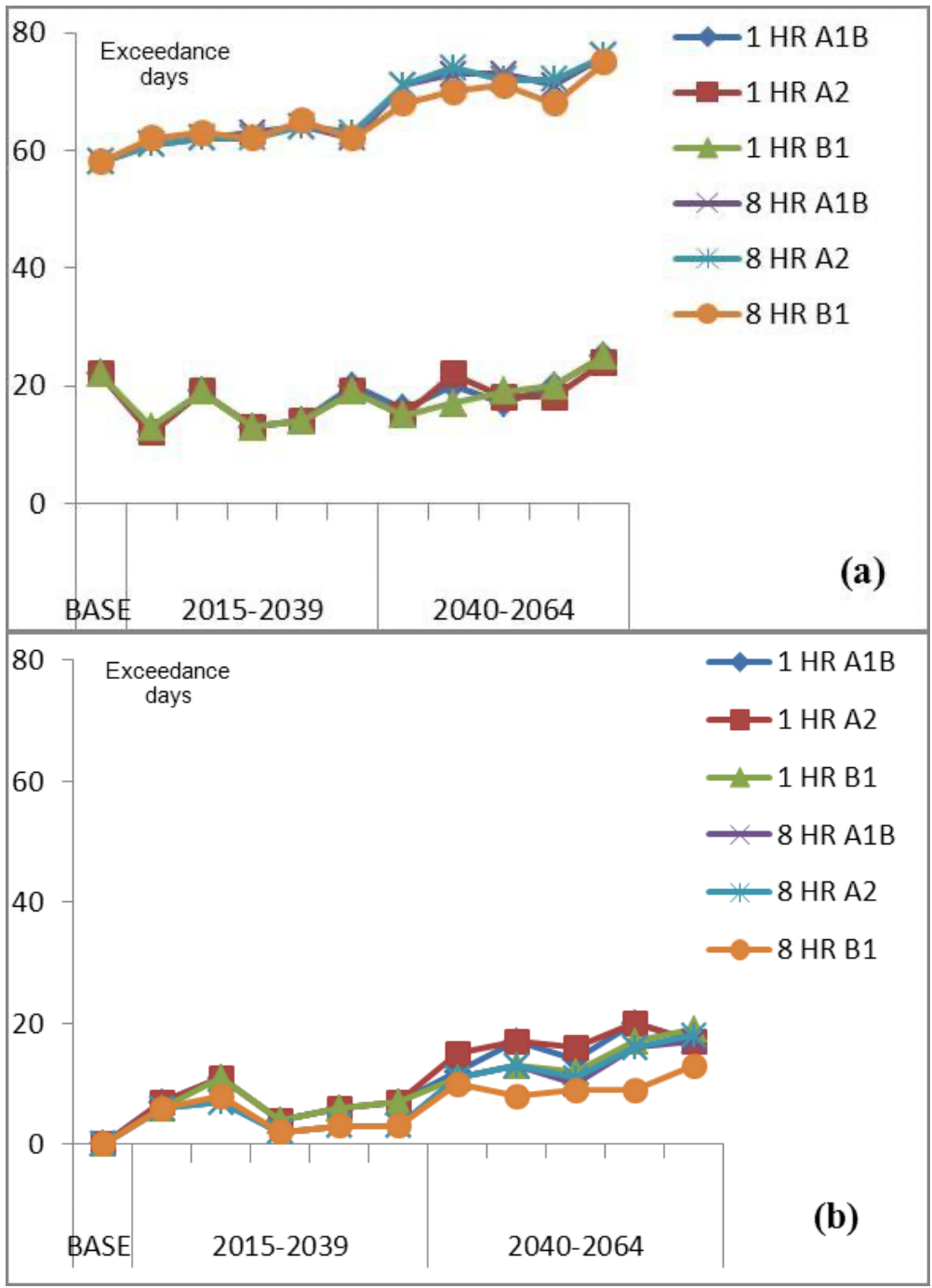

Figure 12. Projected days per summer (JJA) with exceedances of 1-hr and 8-hr O3 stan- dard based on summers of 2010 (a) and 2012 (b).

Regardless of existing uncertainties in different parts of the climate change impact assessment such as uncertainties in climate sensitivity and future greenhouse gasses emission pathways, projections indicate that as a consequence of occurring more favorable O3 formation conditions in the future due to climate change, the number of O3 polluted days will increase over all emission scenarios and climate periods, even based on the violation-free pollution conditions of the summer of 2012. Summer of 2010 was a year with the highest monitored O3 concentrations in the observations probably due to meteorological conditions favorable to O3 formation. About 58 out of 92 days of the 2010 summer violated 8-hr O3 standard. In contrast, 2012 experienced a violation-free summer, which can serve as a suitable representation of present conditions for analyzing the sensitivity of O3 air quality to future climate changes while O3 precursor emissions are held constant over future decades.

Furthermore, comparing changes in the projected O3 exceedances in the two climate periods, short term changes based on each pollution condition are almost overlapped and no noticeable distinction exists between different emission scenarios. However, as Stott and Kettleborough (Stott and Kettleborough, 2002) noted, due to the inertia in the climate system, inter-scenario differences among SRES emission scenarios will emerge after 2030 and the differences between projections are more pronounced in mid-term and long term projections.

### 3.6.2. Projected Average number of Ozone polluted days in each climate period

In the presented projections (Figure 11 and 12), due to the stochastic nature of the downscaling technique, only the changes in the number of exceedances in each climate period should be considered and the projected number of exceedances in a specific summer cannot be associated to its corresponding year in the future. To provide a quantitative comparison between the projected changes in the number of polluted days from the O3 air quality standard perspective, the average number of exceedances in each climate period and emission scenario was calculated and compared.

Figure 13 illustrates the projected average number of exceedances of 8-hr O3 standard under each emission scenario and pollution condition. Based on the pollution conditions in the both summers of 2010 and 2012, the number of polluted days in all scenarios is expected to rise in the future. In the short term, the largest increase in the number of polluted days is anticipated for the B1 emission scenario and in the mid-term the largest increase is projected for the A2 simulations. In the short term and based on the pollution conditions of the summer of 2010, the largest increase is expected to be about 8.3% for the B1 scenario from 58 exceedance days in 2010 to 62.8 exceedance days in the short term. In the mid-term, the largest increase is expected to be about 26% for the A2 scenario from 58 exceedance days in 2010 to 73 exceedance days in the mid-term. Likewise, based on the pollution conditions of the summer of 2012, all scenarios show an increase in the number of exceedance days. In the short term, the largest increase in the number of exceedance days is anticipated for the B1 scenario and this number grows from zero in 2012 to 4.4 days in the short term. In the mid-term, the largest increase in the number of exceedance days is projected for the A2 scenario and this number grows from zero in 2012 to 13.8 days in the mid-term.

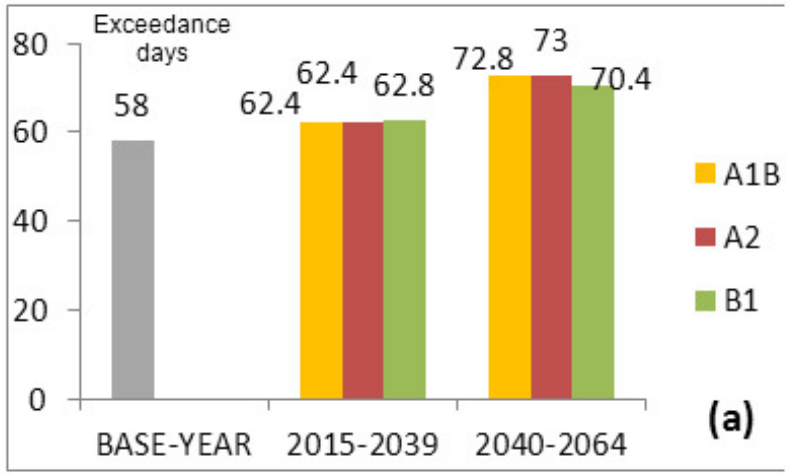

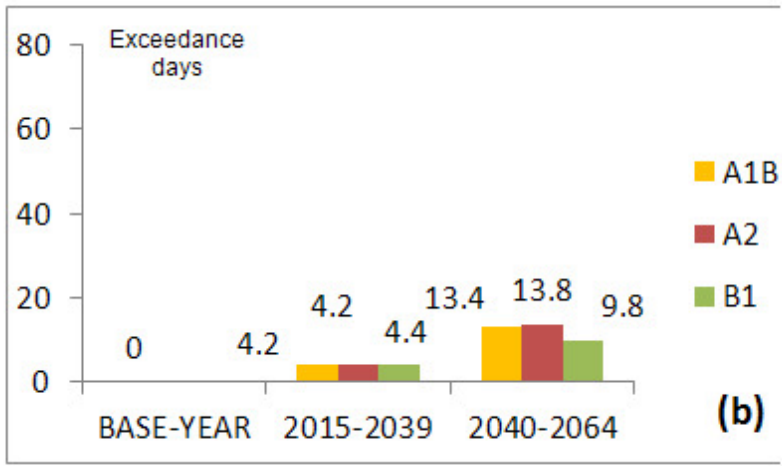

Figure 13. Projected average number of summer days (JJA) with exceedance of 8-hr O3 standard based on summers of 2010 (a) and 2012 (b).

### 3.6.3. Changes in future Ozone Air Quality Index exceedances (AQI)

The projected O3 concentrations were also analyzed from health-related perspectives such as 1-hr and 8-hr O3 Air Quality Indices (AQI). In this section only the 8-hr projections for the A1B emission scenario are presented. Figure 14 illustrates the change in the number of days with exceedance of the 8-hr O3 AQI concentration thresholds under the A1B emission scenario for pollution conditions based on the summers of 2010 (a) and 2012 (b). Based on the both pollution conditions, projections indicate an increase in the number of *Unhealthy* and *Very Unhealthy* days under the impact of climate change which reflects the degradation of O3 air quality in the future.

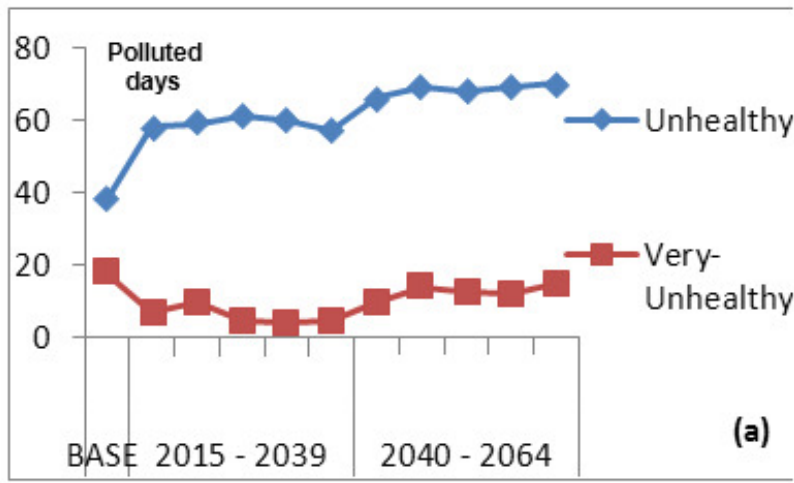

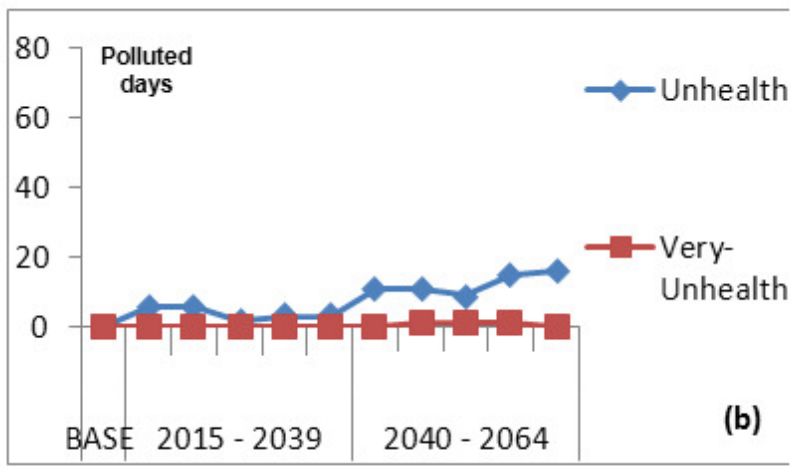

Figure 14. Projected days per summer (JJA) with exceedances of 8-hr O3 AQI based on summers of 2010 (a) and 2012 (b).

Figure 14a indicates that based on pollution conditions of the summer of 2010, the number of *Unhealthy* days is expected to increase over both climate periods, while the number of *Very Unhealthy* days decreases over the first period compared to summer of 2010 and then increases in the second period. Figure 14b indicates that based on pollution conditions of the summer of 2010, the number of *Unhealthy* days grows over the two future climate periods compared to the summer of 2012. Occurrence of the *Very Unhealthy* days are not projected over the first climate period, but due to the projected higher temperature and radiation, the number of the *Very Unhealthy* days starts to grow over the second climate period.

To provide a quantitative comparison between the projected changes in the number of polluted days from AQI perspective, the average number of polluted days was calculated for each climate period. In this section only the projections for the 8-hr O3 AQI under A1B emission scenario are demonstrated. Figure 15 depicts the average number of polluted days from the 8-hr O3 AQI perspective for the pollution conditions in the summers of 2010 (a) and 2012 (b). As the Figure 15a indicates, the average number of *Unhealthy* days is expected to grow over the two future climate periods. The number of *Unhealthy* days was 38 days in the summer of 2010, which is projected to increase about 55% over the first period, averaging about 59 days in the short term, and about 80% over the second period, averaging about 68.5 days in the mid-term, compared to the summer of 2010. Moreover, the number of *Very Unhealthy* days is projected to fall about 65% from 18 days in the summer of 2010 to 6.2 days in the first period, but to double in the second period by increasing from 6.2 days to 12.8 days in the mid-term.

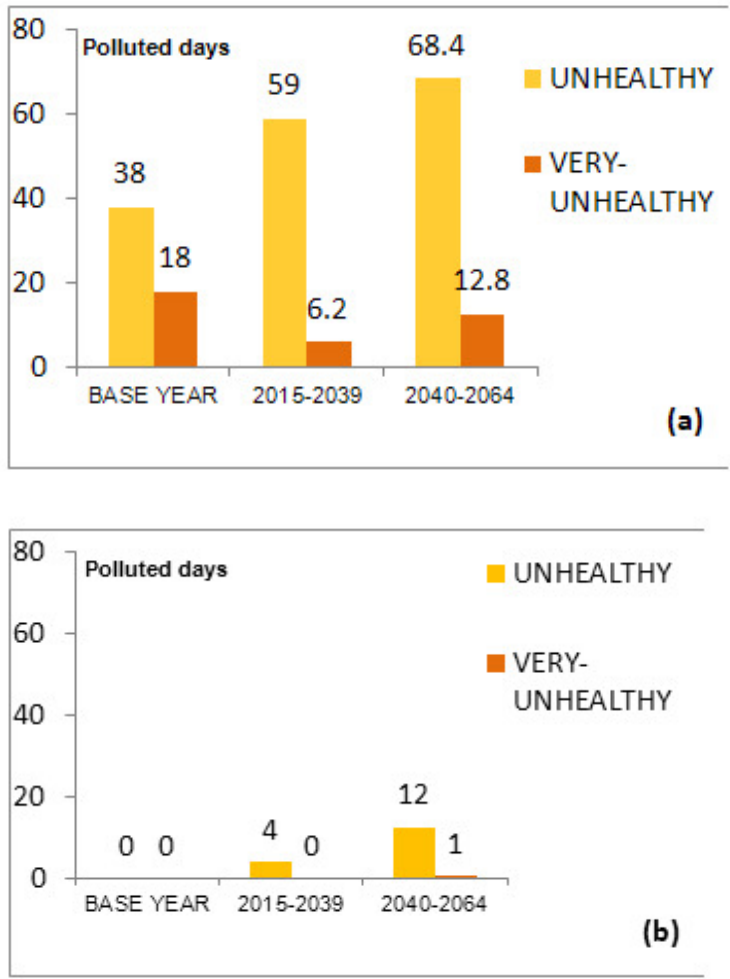

Figure 15. Projected average number of summer days (JJA) with exceedance of 8-hr O3 AQI based on summers of 2010 (a) and 2012 (b).

Similar to Figure 15a, Figure 15b depicts an increase in the average number of polluted days based on the pollution conditions in the summer of 2012. As the graph shows, no *Unhealthy* day occurred in the summer of 2012. However, projections estimate about 4 *Unhealthy* days without any *Very Unhealthy* days in the short term. In the mid-term, the average number of *Unhealthy* days is expected to increase to 12 days with one *Very Unhealthy* day.

4. **Summary and conclusion**

In this study, the impact of climate change on future summertime O3 concentrations in Tehran, Iran was investigated by using a statistical approach. Climate change projections under the IPCC SRES A1B, B1 and A2 emission scenarios over the study area were obtained from HACM3 GCM and were downscaled by LARS-WG5 model over the periods of 2015–2039 and 2040–2064.

Climate change projections indicate that Tehran becomes warmer over the next 50 years. The projected increases in temperature and solar radiation along with the decreases in precipitation and cloud cover for the future summers over the study area are indications of more favorable conditions for photochemical pollution formation which could result in degraded air quality conditions in future summers. To quantify the impact of projected changes in climate variables on future O3 levels, an air quality forecasting model (ANN) was developed based on four predictors of temperature, solar radiation, NO and NO2. The projected summertime hourly O3 concentrations were analyzed based on exceedances of several O3 standards and air quality indices. The projections were conducted by assuming that current emission conditions of O3 precursors remain constant in the future. Therefore, pollution conditions of the summers of 2010 and 2012 were considered as two different pollution scenarios and only the impact of climate change alone were accounted in the projections. The simulations projected that the number of O3 polluted days would increase based on both summer pollution conditions and the increase based on the unpolluted summer of 2012 would be more noticeable compared to the highly polluted summer of 2010. Moreover, the growing number of polluted days in terms of 8-hr indices compared to 1-hr indices could be an indication of more exposure to higher O3 concentrations in the future.

Since this study is considered as the first study in Iran which addresses the influence of future climate on air quality, it was subject to various limitations. One of the major limitations was that NMVOC concentrations were not considered in the simulations due to unavailability of observed data. O3 simulations without considering NMVOCs in the calculation process tend to underestimate O3 concentrations (Liu et al., 1987). Moreover, from several climate variables effective in O3 production, due to simplification in the modeling process, only temperature and solar radiation were selected although O3 formation is sensitive to other climate variables such as wind speed, water vapor, cloud cover or precipitation (Dawson et al., 2007). Another imposed limitation, which is associated with statistical approaches, is the assumption that emissions of O3 precursors and their relationship with O3 formation remain constant in the future and therefore the role of future emission reductions cannot be considered in the simulations. To reduce the scope of this limitation in simulations, two different summertime pollution conditions with the highest and the lowest number of monitored polluted days were considered in calculations to demonstrate the probable range of changes in O3 simulations in each pollution condition. However, a number of sensitivity studies have suggested that climate change may offset benefits of anticipated future emissions reductions in some regions and additional supervision on present emission reduction policies is recommended (Millstein and Harley, 2009; Steiner et al., 2006).

Future research should therefore concentrate on the limitations of this study. Since the absence of NMVOC concentrations as one of the main precursors of O3 production reduces the accuracy in the simulations, future studies should benefit from including NMVOC concentrations in the simulation process. Moreover, with regard to existing uncertainties in GCM projections, future studies should also benefit from ensemble projection approaches by incorporating several GCMs in climate change impact assessments to improve the scope of confidence in air quality projections.

Furthermore, using dynamical downscaling results from Regional Climate Models, including other climate variables in projections, and comparing projections of statistical approaches with projections of Chemistry Transport Models could be other beneficial measures for improving the accuracy and confidence in the climate change impact assessments.

**References**
Alibak, A.H., Khodarahmi, M., Fayyazsanavi, P., Alizadeh, S.M., Hadi, A.J., Aminzadehsarikhanbeglou, E., 2022. Simulation the adsorption capacity of polyvinyl alcohol/carboxymethyl cellulose based hydrogels towards methylene blue in aqueous solutions using cascade correlation neural network (CCNN) technique. J. Clean. Prod. 337, 130509. https://doi.org/10.1016/j.jclepro.2022.130509

Arhami, M., Kamali, N., Rajabi, M.M., 2013. Predicting hourly air pollutant levels using artificial neural networks coupled with uncertainty analysis by Monte Carlo simulations. Environ. Sci. Pollut. Res. 20, 4777–4789. https://doi.org/10.1007/s11356-012-1451-6

Ashrafi, K., 2012. Determining of spatial distribution patterns and temporal trends of an air pollutant using proper orthogonal decomposition basis functions. Atmos. Environ. 47, 468–476. https://doi.org/10.1016/j.atmosenv.2011.10.016

Atash, F., 2007. The deterioration of urban environments in developing countries: Mitigating the air pollution crisis in Tehran, Iran. Cities 24, 399–409. https://doi.org/10.1016/j.cities.2007.04.001

Baertsch-Ritter, N., Keller, J., Dommen, J., Prevot, A.S.H., 2004. Effects of various meteorological conditions and spatial emissionresolutions on the ozone concentration and ROG/$NO_x$ limitationin the Milan area (I). Atmospheric Chem. Phys. 4, 423–438. https://doi.org/10.5194/acp-4-423-2004

Beale, M.H., Hagan, M.T., Demuth, H.B., 2012. Neural Network Toolbox™ User's Guide, in: R2012a, The MathWorks, Inc., 3 Apple Hill Drive Natick, MA 01760-2098, , Www.Mathworks.Com.

Bell, M.L., Goldberg, R., Hogrefe, C., Kinney, P.L., Knowlton, K., Lynn, B., Rosenthal, J., Rosenzweig, C., Patz, J.A., 2007. Climate change, ambient ozone, and health in 50 US cities. Clim. Change 82, 61–76. https://doi.org/10.1007/s10584-006-9166-7

Bell, M.L., McDermott, A., Zeger, S.L., Samet, J.M., Dominici, F., 2004. Ozone and Short-term Mortality in 95 US Urban Communities, 1987-2000. JAMA 292, 2372–2378. https://doi.org/10.1001/jama.292.19.2372

Camalier, L., Cox, W., Dolwick, P., 2007. The effects of meteorology on ozone in urban areas and their use in assessing ozone trends. Atmos. Environ. 41, 7127–7137. https://doi.org/10.1016/j.atmosenv.2007.04.061

Chaloulakou, A., Saisana, M., Spyrellis, N., 2003. Comparative assessment of neural networks and regression models for forecasting summertime ozone in Athens. Sci. Total Environ. 313, 1–13. https://doi.org/10.1016/S0048-9697(03)00335-8

Comrie, A.C., 1997. Comparing Neural Networks and Regression Models for Ozone Forecasting. J. Air Waste Manag. Assoc. 47, 653–663. https://doi.org/10.1080/10473289.1997.10463925

Cox, W.M., Chu, S.-H., 1996. Assessment of interannual ozone variation in urban areas from a climatological perspective. Atmos. Environ. 30, 2615–2625. https://doi.org/10.1016/1352-2310(95)00346-0

Dawson, J.P., Adams, P.J., Pandis, S.N., 2007. Sensitivity of ozone to summertime climate in the eastern USA: A modeling case study. Atmos. Environ. 41, 1494–1511. https://doi.org/10.1016/j.atmosenv.2006.10.033

Dawson, J.P., Racherla, P.N., Lynn, B.H., Adams, P.J., Pandis, S.N., 2009. Impacts of climate change on regional and urban air quality in the eastern United States: Role of meteorology. J. Geophys. Res. Atmospheres 114. https://doi.org/10.1029/2008JD009849

Ebi, K.L., McGregor, G., 2008. Climate Change, Tropospheric Ozone and Particulate Matter, and Health Impacts. Environ. Health Perspect. 116, 1449–1455. https://doi.org/10.1289/ehp.11463

Ephrath, J.E., Goudriaan, J., Marani, A., 1996. Modelling diurnal patterns of air temperature, radiation wind speed and relative humidity by equations from daily characteristics. Agric. Syst. 51, 377–393. https://doi.org/10.1016/0308-521X(95)00068-G

Fuentes, J.D., Lerdau, M., Atkinson, R., Baldocchi, D., Bottenheim, J.W., Ciccioli, P., Lamb, B., Geron, C., Gu, L., Guenther, A., Sharkey, T.D., Stockwell, W., 2000. Biogenic Hydrocarbons in the Atmospheric Boundary Layer: A Review. Bull. Am. Meteorol. Soc. 81, 1537–1575.

Gardner, M.W., Dorling, S.R., 2000. Statistical surface ozone models: an improved methodology to account for non-linear behaviour. Atmos. Environ. 34, 21–34. https://doi.org/10.1016/S1352-2310(99)00359-3

Gardner, M.W., Dorling, S.R., 1998. Artificial neural networks (the multilayer perceptron)—a review of applications in the atmospheric sciences. Atmos. Environ. 32, 2627–2636. https://doi.org/10.1016/S1352-2310(97)00447-0

Gryparis, A., Forsberg, B., Katsouyanni, K., Analitis, A., Touloumi, G., Schwartz, J., Samoli, E., Medina, S., Anderson, H.R., Niciu, E.M., Wichmann, H.-E., Kriz, B., Kosnik, M., Skorkovsky, J., Vonk, J.M., Dörtbudak, Z., 2004. Acute Effects of Ozone on Mortality from the "Air Pollution and Health. Am. J. Respir. Crit. Care Med. 170, 1080–1087. https://doi.org/10.1164/rccm.200403-333OC

Guenther, A., Geron, C., Pierce, T., Lamb, B., Harley, P., Fall, R., 2000. Natural emissions of non-methane volatile organic compounds, carbon monoxide, and oxides of nitrogen from North America. Atmos. Environ. 34, 2205–2230. https://doi.org/10.1016/S1352-2310(99)00465-3

Halek, F., Kavouci, A., Montehaie, H., 2004. Role of motor-vehicles and trend of air borne particulate in the Great Tehran area, Iran. Int. J. Environ. Health Res. 14, 307–313. https://doi.org/10.1080/09603120410001725649

Hessami, M., Gachon, P., Ouarda, T.B.M.J., St-Hilaire, A., 2008. Automated regression-based statistical downscaling tool. Environ. Model. Softw. 23, 813–834. https://doi.org/10.1016/j.envsoft.2007.10.004

Hogrefe, C, Biswas, J., Lynn, B., Civerolo, K., Ku, J.-Y., Rosenthal, J., Rosenzweig, C., Goldberg, R., Kinney, P.L., 2004. Simulating regional-scale ozone climatology over the eastern United States: model evaluation results. Atmos. Environ. 38, 2627–2638. https://doi.org/10.1016/j.atmosenv.2004.02.033

Hogrefe, C., Lynn, B., Civerolo, K., Ku, J.-Y., Rosenthal, J., Rosenzweig, C., Goldberg, R., Gaffin, S., Knowlton, K., Kinney, P.L., 2004. Simulating changes in regional air pollution over the eastern United States due to changes in global and regional climate and emissions. J. Geophys. Res. Atmospheres 109. https://doi.org/10.1029/2004JD004690

Holloway, T., Spak, S.N., Barker, D., Bretl, M., Moberg, C., Hayhoe, K., Van Dorn, J., Wuebbles, D., 2008. Change in ozone air pollution over Chicago associated with global climate change. J. Geophys. Res. Atmospheres 113. https://doi.org/10.1029/2007JD009775

Hosseinpoor, A.R., Forouzanfar, M.H., Yunesian, M., Asghari, F., Naieni, K.H., Farhood, D., 2005. Air pollution and hospitalization due to angina pectoris in Tehran, Iran: A

time-series study. Environ. Res. 99, 126–131. https://doi.org/10.1016/j.envres.2004.12.004

Hoveidi, H., Aslemand, A., Vahidi, H., Akhavan, F., 2013. Cost Emission of Pm10 on Human Health Due to the Solid Waste Disposal Scenarios, Case Study; Tehran, Iran. J. Earth Sci. Clim. Change. https://doi.org/10.4172/2157-7617.1000139

IPCC, 2007: Summary for Policymakers. In: Climate Change 2007: The Physical Science Basis. Contribution of Working Group I to the Fourth Assessment Report of the Intergovernmental Panel on Climate Change [Solomon, S., D. Qin, M. Manning, Z. Chen, M. Marquis, K.B. Averyt, M.Tignor and H.L. Miller (eds.)]. Cambridge University Press, Cambridge, United Kingdom and New York, NY, USA.

IPCC, 2021: Summary for Policymakers. In: Climate Change 2021: The Physical Science Basis. Contribution of Working Group I to the Sixth Assessment Report of the Intergovernmental Panel on Climate Change [Masson-Delmotte, V., P. Zhai, A. Pirani, S.L. Connors, C. Péan, S. Berger, N. Caud, Y. Chen, L. Goldfarb, M.I. Gomis, M. Huang, K. Leitzell, E. Lonnoy, J.B.R. Matthews, T.K. Maycock, T. Waterfield, O. Yelekçi, R. Yu, and B. Zhou (eds.)]. Cambridge University Press, Cambridge, United Kingdom and New York, NY, USA, pp. 3–32, doi:10.1017/9781009157896.001.

Ito, K., De Leon, S.F., Lippmann, M., 2005. Associations between Ozone and Daily Mortality: Analysis and Meta-Analysis. Epidemiology 16, 446–457.

Jacob, D.J., Logan, J.A., Gardner, G.M., Yevich, R.M., Spivakovsky, C.M., Wofsy, S.C., Sillman, S., Prather, M.J., 1993. Factors regulating ozone over the United States and its export to the global atmosphere. J. Geophys. Res. Atmospheres 98, 14817–14826. https://doi.org/10.1029/98JD01224

Jacob, D.J., Winner, D.A., 2009. Effect of climate change on air quality. Atmos. Environ., Atmospheric Environment - Fifty Years of Endeavour 43, 51–63. https://doi.org/10.1016/j.atmosenv.2008.09.051

Jacobson, M.Z., 2005. Fundamentals of Atmospheric Modeling, 2nd ed. Cambridge University Press, Cambridge. https://doi.org/10.1017/CBO9781139165389

Lee, B.-S., Wang, J.-L., 2006. Concentration variation of isoprene and its implications for peak ozone concentration. Atmos. Environ. 40, 5486–5495. https://doi.org/10.1016/j.atmosenv.2006.03.035

Leibensperger, E.M., Mickley, L.J., Jacob, D.J., 2008. Sensitivity of US air quality to mid-latitude cyclone frequency and implications of 1980–2006 climate change. Atmospheric Chem. Phys. 8, 7075–7086. https://doi.org/10.5194/acp-8-7075-2008

Liao, H., Chen, W.-T., Seinfeld, J.H., 2006. Role of climate change in global predictions of future tropospheric ozone and aerosols. J. Geophys. Res. Atmospheres 111. https://doi.org/10.1029/2005JD006852

Lioubimtseva, E., Henebry, G.M., 2009. Climate and environmental change in arid Central Asia: Impacts, vulnerability, and adaptations. J. Arid Environ. 73, 963–977. https://doi.org/10.1016/j.jaridenv.2009.04.022

Liu, S.C., Trainer, M., Fehsenfeld, F.C., Parrish, D.D., Williams, E.J., Fahey, D.W., Hübler, G., Murphy, P.C., 1987. Ozone production in the rural troposphere and the implications for regional and global ozone distributions. J. Geophys. Res. Atmospheres 92, 4191–4207. https://doi.org/10.1029/JD092iD04p04191

Lynn, B.H., Druyan, L., Hogrefe, C., Dudhia, J., Rosenzweig, C., Goldberg, R., Rind, D., Healy, R., Rosenthal, J., Kinney, P., 2004. Sensitivity of present and future surface temperatures to precipitation characteristics. Clim. Res. 28, 53–65. https://doi.org/10.3354/cr028053
Marzban, C., Stumpf, G., 1996. A Neural Network for Tornado Prediction Based on Doppler Radar-Derived Attributes. https://doi.org/10.1175/1520-0450(1996)035<0617:ANNFTP>2.0.CO;2
Mejia, J., Wilcox, E., Rayne, S., Mosadegh, E., 2018. Final report: Vehicle Miles Traveled Review. https://doi.org/10.13140/RG.2.2.29814.52807
Mickley, L.J., Jacob, D.J., Field, B., Rind, D.M., 2004. Effects of future climate change on regional air pollution episodes in the United States. Geophys. Res. Lett. https://doi.org/10.1029/2004GL021216
Millstein, D.E., Harley, R.A., 2009. Impact of climate change on photochemical air pollution in Southern California. Atmospheric Chem. Phys. 9, 3745–3754. https://doi.org/10.5194/acp-9-3745-2009
Mosadegh, E., 2013. Modeling the Regional Effects of Climate Change on Future Urban Air Quality (With Special Reference to Future Ozone Concentrations in Tehran, Iran). Univ. Tehran, Iran. DOI: 10.13140/RG.2.2.23815.32165
Mosadegh, E., Babaeian, I., 2022a. Projection of Temperature and Precipitation for 2020-2100 Using Post-processing of General Circulation Models Output and Artificial Neural Network Approach, Case Study: Tehran and Alborz Provinces. Iranian Journal of Geophysics. https://doi.org/10.30499/ijg.2022.311104.1370
Mosadegh, E., Babaeian, I., 2022b. Quantifying the Range of Uncertainty in GCM Projections for Future Solar Radiation, Temperature and Precipitation under Global Warming Effect in Dushan-Tappeh Station, Tehran, Iran. Iranian Journal of Geophysics. https://doi.org/10.30499/ijg.2022.310958.1369
Mosadegh, E., Mejia, J., Wilcox, E.M., Rayne, S., 2018. Vehicle Miles Travel (VMT) trends over Lake Tahoe area and its effect on Nitrogen Deposition 2018, A23M-3068.
Mosadegh, E., Nolin, A.W., 2020. Estimating Arctic sea ice surface roughness by using back propagation neural network 2020, C014-0005.
Mott, J.A., Mannino, D.M., Alverson, C.J., Kiyu, A., Hashim, J., Lee, T., Falter, K., Redd, S.C., 2005. Cardiorespiratory hospitalizations associated with smoke exposure during the 1997, Southeast Asian forest fires. Int. J. Hyg. Environ. Health 208, 75–85. https://doi.org/10.1016/j.ijheh.2005.01.018
Mudway, I., Kelly, F., 2000. Ozone and the lung: A sensitive issue. Mol. Aspects Med. 21, 1–48. https://doi.org/10.1016/S0098-2997(00)00003-0
Murazaki, K., Hess, P., 2006. How does climate change contribute to surface ozone change over the United States? J. Geophys. Res. Atmospheres 111. https://doi.org/10.1029/2005JD005873
Narumi, D., Kondo, A., Shimoda, Y., 2009. The effect of the increase in urban temperature on the concentration of photochemical oxidants. Atmos. Environ. 43, 2348–2359. https://doi.org/10.1016/j.atmosenv.2009.01.028
Nejatishahidin, N., Fayyazsanavi, P., Kosecka, J., 2022. Object Pose Estimation using Mid-level Visual Representations (No. arXiv:2203.01449). arXiv. https://doi.org/10.48550/arXiv.2203.01449
Nilsson, E.D., Paatero, J., Boy, M., 2001a. Effects of air masses and synoptic weather on aerosol formation in the continental boundary layer. Tellus B 53, 462–478. https://doi.org/10.1034/j.1600-0889.2001.530410.x
37

Nilsson, E.D., Rannik, Ü., Kumala, M., Buzorius, G., O'dowd, C.D., 2001b. Effects of continental boundary layer evolution, convection, turbulence and entrainment, on aerosol formation. Tellus B Chem. Phys. Meteorol. 53, 441–461. https://doi.org/10.3402/tellusb.v53i4.16617

Niska, H., Hiltunen, T., Karppinen, A., Ruuskanen, J., Kolehmainen, M., 2004. Evolving the neural network model for forecasting air pollution time series. Eng. Appl. Artif. Intell., Intelligent Control and Signal Processing 17, 159–167. https://doi.org/10.1016/j.engappai.2004.02.002

Noori, R., Hoshyaripour, G., Ashrafi, K., Araabi, B.N., 2010. Uncertainty analysis of developed ANN and ANFIS models in prediction of carbon monoxide daily concentration. Atmos. Environ. 44, 476–482. https://doi.org/10.1016/j.atmosenv.2009.11.005

Nunnari, G., Nucifora, A.F.M., Randieri, C., 1998. The application of neural techniques to the modelling of time-series of atmospheric pollution data. Ecol. Model. 111, 187–205. https://doi.org/10.1016/S0304-3800(98)00118-5

Ordóñez, C., Mathis, H., Furger, M., Henne, S., Hüglin, C., Staehelin, J., Prévôt, A.S.H., 2005. Changes of daily surface ozone maxima in Switzerland in all seasons from 1992 to 2002 and discussion of summer 2003. Atmospheric Chem. Phys. 5, 1187–1203. https://doi.org/10.5194/acp-5-1187-2005

Orru, H., Andersson, C., Ebi, K.L., Langner, J., Åström, C., Forsberg, B., 2013. Impact of climate change on ozone-related mortality and morbidity in Europe. Eur. Respir. J. 41, 285–294. https://doi.org/10.1183/09031936.00210411

Racherla, P.N., Adams, P.J., 2006. Sensitivity of global tropospheric ozone and fine particulate matter concentrations to climate change. J. Geophys. Res. Atmospheres 111. https://doi.org/10.1029/2005JD006939

Rahnama, M., Noury, M., 2008. Developing of Halil River Rainfall-Runoff Model, Using Conjunction of Wavelet Transform and Artificial Neural Networks. Res. J. Environ. Sci. 2, 385–392. https://doi.org/10.3923/rjes.2008.385.392

Schlink, U., Dorling, S., Pelikan, E., Nunnari, G., Cawley, G., Junninen, H., Greig, A., Foxall, R., Eben, K., Chatterton, T., Vondracek, J., Richter, M., Dostal, M., Bertucco, L., Kolehmainen, M., Doyle, M., 2003. A rigorous inter-comparison of ground-level ozone predictions. Atmos. Environ. 37, 3237–3253. https://doi.org/10.1016/S1352-2310(03)00330-3

Seinfeld, J.H., Pandis, S.N., 2006. Atmospheric chemistry and physics: from air pollution to climate change, 2nd ed. ed. Wiley, Hoboken, N.J.

Semenov, M., Barrow, E., 2002. LARS-WG A Stochastic Weather Generator for Use in Climate Impact Studies.

Semenov, M., Stratonovitch, P., 2010. Use of multi-model ensembles from global climate models for assessment of climate change impacts. Clim. Res. - Clim. RES 41, 1–14. https://doi.org/10.3354/cr00836

Semenov, M.A., 2007. Development of high-resolution UKCIP02-based climate change scenarios in the UK. Agric. For. Meteorol. 144, 127–138. https://doi.org/10.1016/j.agrformet.2007.02.003

Sillman, S., 1999. The relation between ozone, NOx and hydrocarbons in urban and polluted rural environments. Atmos. Environ. 33, 1821–1845. https://doi.org/10.1016/S1352-2310(98)00345-8

Sillman, S., Samson, P.J., 1995. Impact of temperature on oxidant photochemistry in urban, polluted rural and remote environments. J. Geophys. Res. Atmospheres 100, 11497–11508. https://doi.org/10.1029/94JD02146

Sousa, S.I.V., Martins, F.G., Alvim-Ferraz, M.C.M., Pereira, M.C., 2007. Multiple linear regression and artificial neural networks based on principal components to predict ozone concentrations. Environ. Model. Softw. 22, 97–103. https://doi.org/10.1016/j.envsoft.2005.12.002

Sousounis, P.J., Scott, C.P.J., Wilson, M.L., 2002. Possible Climate Change Impacts on Ozone in the Great Lakes Region: Some Implications for Respiratory Illness. J. Gt. Lakes Res. 28, 626–642. https://doi.org/10.1016/S0380-1330(02)70610-2

Spitters, C.J.T., Toussaint, H.A.J.M., Goudriaan, J., 1986. Separating the diffuse and direct component of global radiation and its implications for modeling canopy photosynthesis Part I. Components of incoming radiation. Agric. For. Meteorol. 38, 217–229. https://doi.org/10.1016/0168-1923(86)90060-2

Steiner, A.L., Davis, A.J., Sillman, S., Owen, R.C., Michalak, A.M., Fiore, A.M., 2010. Observed suppression of ozone formation at extremely high temperatures due to chemical and biophysical feedbacks. Proc. Natl. Acad. Sci. 107, 19685–19690. https://doi.org/10.1073/pnas.1008336107

Steiner, A.L., Tonse, S., Cohen, R.C., Goldstein, A.H., Harley, R.A., 2006. Influence of future climate and emissions on regional air quality in California. J. Geophys. Res. Atmospheres 111. https://doi.org/10.1029/2005JD006935

Stott, P.A., Kettleborough, J.A., 2002. Origins and estimates of uncertainty in predictions of twenty-first century temperature rise. Nature 416, 723–726. https://doi.org/10.1038/416723a

Syri, S., Karvosenoja, N., Lehtilä, A., Laurila, T., Lindfors, V., Tuovinen, J.-P., 2002. Modeling the impacts of the Finnish Climate Strategy on air pollution. Atmos. Environ. 36, 3059–3069. https://doi.org/10.1016/S1352-2310(02)00263-7

Varotsos, K.V., Tombrou, M., Giannakopoulos, C., 2013. Statistical estimations of the number of future ozone exceedances due to climate change in Europe. J. Geophys. Res. Atmospheres 118, 6080–6099. https://doi.org/10.1002/jgrd.50451

Webster, M.D., Babiker, M., Mayer, M., Reilly, J.M., Harnisch, J., Hyman, R., Sarofim, M.C., Wang, C., 2002. Uncertainty in emissions projections for climate models. Atmos. Environ. 36, 3659–3670. https://doi.org/10.1016/S1352-2310(02)00245-5

Wilby, R., Charles, S., Zorita, E., Timbal, B., Whetton, P., Mearns, L., 2004. Guidelines For Use of Climate Scenarios Developed From Statistical Downscaling Methods. Support. Mater. Intergov. Penel Clim. Change.

Wilks, D.S., Wilby, R.L., 1999. The weather generation game: a review of stochastic weather models. Prog. Phys. Geogr. Earth Environ. 23, 329–357. https://doi.org/10.1177/030913339902300302

Wise, E.K., 2009. Climate-based sensitivity of air quality to climate change scenarios for the southwestern United States. Int. J. Climatol. 29, 87–97. https://doi.org/10.1002/joc.1713

Wise, E.K., Comrie, A.C., 2005. Meteorologically adjusted urban air quality trends in the Southwestern United States. Atmos. Environ. 39, 2969–2980. https://doi.org/10.1016/j.atmosenv.2005.01.024

Zarghami, M., Abdi, A., Babaeian, I., Hassanzadeh, Y., Kanani, R., 2011. Impacts of climate change on runoffs in East Azerbaijan, Iran. Glob. Planet. Change 78, 137–146. https://doi.org/10.1016/j.gloplacha.2011.06.003